\documentclass{aa}

\usepackage{graphicx}
\usepackage{xcolor}
\usepackage{txfonts}
\usepackage{hyperref}

\def\msun{{\rm ~M}_{\odot}}
\def\rsun{{\rm ~R}_{\odot}}

\def\gpy{{\rm ~Gpc}^{-3} {\rm ~yr}^{-1}}
\def\kms{{\rm ~km} {\rm ~s}^{-1}}

\begin{document}

   \title{Black hole--neutron star mergers: The first mass gap and kilonovae}

   \titlerunning{BH-NS mergers and the first mass gap}

   \author{P. Drozda\inst{1}\thanks{paweldro1996@gmail.com}
          \and
          K. Belczynski\inst{2}
          \and
          R.~O'Shaughnessy\inst{3}
          \and
          T. Bulik\inst{1}
          \and
          C.L. Fryer\inst{4}
   }

   \institute{Astronomical Observatory, Warsaw University, Al. Ujazdowskie 4,
           00-478 Warsaw, Poland
         \and
           Nicolaus Copernicus Astronomical Center, Polish Academy of Sciences,
           ul. Bartycka 18, 00-716 Warsaw, Poland
         \and
           Center for Computational Relativity and Gravitation, Rochester Institute of
           Technology, Rochester, NY 14623, USA
         \and
           CCS-2, MSD409, Los Alamos National Laboratory, Los Alamos, NM 87545, USA
   }

  \abstract {
Observations of X-ray binaries indicate a dearth of compact objects in the 
mass range from $\sim 2-5\msun$.  The existence of this (first mass) gap 
has been used to discriminate between proposed engines behind core-collapse 
supernovae.  From LIGO/Virgo observations of binary
compact remnant masses,  several candidate first mass gap objects (either neutron stars (NSs) 
or black holes (BHs)) were identified during the O3 science run.
Motivated by these new observations, we study the formation of BH-NS mergers in the 
framework of isolated classical binary evolution, using  population synthesis methods 
to evolve large populations of binary stars 
(Population I and II) across cosmic time. 
We present results on the NS to BH mass ratios ($q=M_{\rm NS}/M_{\rm BH}$ ) 
in merging systems, showing that although systems with a mass ratio as low as 
$q=0.02$ can exist, typically BH-NS systems form with moderate mass ratios 
$q=0.1-0.2$. 
If we adopt a delayed supernova engine, we conclude that $\sim 30\%$ of BH-NS mergers may 
host at least one compact object in the first mass gap (FMG$^\circ$). 
Even allowing for uncertainties in the processes behind compact object formation, we expect 
the fraction  of BH-NS systems ejecting mass during the merger to be small (from 
$\sim 0.6-9\%$).  
In our reference model, we assume: (i) the formation of compact objects within 
the FMG, (ii) natal NS/BH kicks decreased by fallback, (iii) low BH 
spins due to Tayler-Spruit angular momentum transport in massive stars. We find 
that  $\lesssim 1\%$ of BH-NS mergers will have any mass ejection and about 
the same percentage will produce kilonova bright enough to have a chance of being detected with a large (Subaru-class) $8$m telescope. Interestingly, all 
these mergers will have both a BH and an NS in the FMG.
}


\maketitle

\section{Introduction}
\label{intro}

Detailed analyses of X-ray binary observations have found a paucity of
compact remnants, neutron stars (NSs) and black holes (BHs), in a mass
range from $\sim 2-5\msun$ 
(\cite{bailyn1998mass},\cite{ozel2010black}).  These observations
contradicted the prevailing theory at the time that argued for a
continuous distribution of compact remnants across this
range~\citep{fryer01}.  One explanation for this mass gap is that the
current systems were the result of an observational
bias~\citep{farr2011mass,Kreidberg2012}.  Certainly, there has been a
set of observed systems with measurements within the mass gap
region, for example: the recent observation of an NS with a mass
of $2.14^{+0.10}_{-0.09}\msun$\citep{cromartie2019very}, the
low-mass X-ray binary with a black hole mass of $3.3^{+2.8}_{-0.7}\msun$ 
(\cite{thompson2019noninteracting}), and the ambiguous smaller component in the 
merger event GW190814 with mass $m_2=2.59^{+0.08}_{-0.09}\msun$ \citep{gw190814}.  
It was argued that the existence, or nonexistence, of compact objects in this gap 
region could be used to constrain the properties of the core-collapse supernova engine 
~\citep{belczynski2012missing}. Alternatively, compact objects in the mass gap could 
be created by mergers of lighter compact objects, and thus probe globular cluster 
dynamics. This would allow such compact objects to acquire companions, which leads to
second generation merges \citep{fragione2019black,fragione2020merging}. 

So far, no BH-NS binaries have been detected through electromagnetic (EM) observations 
\citep{2014MNRAS.445.3115L,bhattacharya2019mergers}. Although observations of X-ray 
binaries suggest that such systems should exist, the formation rate of merging BH-NS 
binaries is not well constrained through EM observations~\citep{Belczynski2011,
Belczynski2012,Belczynski2013,Beldycki2016}. 

The LIGO/Virgo gravitational wave detectors are ideally suited to studying the properties
of compact remnants and answering questions about the existence of a mass
gap. The primary gravitational wave (GW) signals for this ground-based
detector consortium arise from the merger of double compact objects:
BH-BH, BH-NS, and NS-NS. The first two LIGO/Virgo science runs (O1 and O2)  
detected 11 merging compact binaries: 10 BH-BH and 1 NS-NS~\citep{abbott2019gwtc}.  
Although these observations brought a wealth of information about BHs and NSs,  
no BH-NS mergers were observed and no compact object was reported to have mass in 
the mass gap region~\citep{LIGO2019a}.

This picture is changing as data from the third (O3) LIGO/Virgo science run are published. 
Table~\ref{tab.events} shows the most confident candidates associated with the mass 
gap from the O3 science run. Four events were classified as BH-NS mergers
~\citep{bhns2021,LIGO2021c}, and one as either a BH-BH or a BH-NS merger because the lighter 
compact object lies in the first mass gap (FMG) and its nature is currently unknown~\citep{gw190814}.

We note that LIGO/Virgo's prompt classification scheme adopts specific choices about how 
to characterize compact objects that differ  from those typically assumed
in the literature: NSs are objects with mass 
$M<3\msun$, FMG objects have mass $3<M<5\msun$, and BHs have mass $M>5\msun$.  This 
classification means that any type of actual double compact object (NS-NS, BH-NS, or 
BH-BH) can have one or two FMG objects as long as the true mass limit discriminating 
between  NSs and BHs is anywhere between $\sim 2-5\msun$. 

GW (gravitational wave) ground based observations can identify the FMG objects, but they cannot easily distinguish between low-mass BHs or high-mass NSs (e.g., \cite{Hinderer2019}). 
EM observations can possibly help to identify the nature of a given FMG object. For 
example, the expectation is that BH-BH mergers do not produce EM emission, and no EM 
counterpart has been found so far for such a merger (e.g., \cite{Greiner2016}).  For BH-NS and 
NS-NS mergers, various EM counterparts are expected across the EM spectrum. The most notable 
are short gamma-ray bursts~\citep{paczynski91,Lee1995,Ruffert1997} and kilonovae
\citep{li1998transient}.  Although there are expected differences between the composition 
of the ejecta, depending on the size and nature (NS vs. BH) of the compact remnant, the 
differences are currently difficult to disentangle from the many model uncertainties (e.g.,
\cite{Metzger2017,miller19,korobkin20}). However, by better understanding BH-NS properties, 
there is hope to better differentiate between BH-NS and NS-NS systems.

In this study we focus on BH-NS mergers that form in classical (common envelope (CE))
isolated 
binary evolution (e.g., \cite{Belczynski2016b}). We estimate the typical physical properties 
of these mergers that  LIGO/Virgo could detect. Using population synthesis models that include 
updated estimates of BH spins and masses \citep{belczynski2019evolutionary}, we estimate 
the fraction of BH-NS mergers that host FMG objects. We also estimate the amount of mass 
ejection during BH-NS mergers, estimating the nature and detectability of any associated kilonova emission.

\begin{table}[b]
\caption{LIGO/Virgo BH-NS merger candidates}
\centering
\begin {tabular}{ l c c c l}
\hline\hline
ID & ${\rm m}_1/\msun$ & ${\rm m}_2/\msun$ & $q$ & type\\
\hline\hline

GW190426 &  $5.7^{+3.9}_{-2.3}$ &    $1.5^{+0.8}_{-0.5}$ & $0.27^{+0.42}_{-0.17}$ & BH-NS \\
         &&&&\\
GW190814 & $23.2^{+1.1}_{-1.0}$ & $2.59^{+0.08}_{-0.09}$ & $0.112^{+0.008}_{-0.009}$ & BH-FMG$^{a}$ \\ 
         &&&&\\
GW190917 &  $9.3^{+3.4}_{-4.4}$ &    $2.1^{+1.5}_{-0.5}$ &     $0.2^{+0.34}_{-0.09}$ & BH-NS \\
         &&&&\\
GW200105 &  $8.9^{+1.1}_{-1.3}$ &    $1.9^{+0.2}_{-0.2}$ &    $0.21^{+0.06}_{-0.04}$ & BH-NS \\
         &&&&\\
GW200115 &  $5.9^{+1.4}_{-2.1}$ &    $1.4^{+0.6}_{-0.2}$ &    $0.24^{+0.31}_{-0.08}$ & BH-NS \\

\hline\hline
\end{tabular}
$^{a}$: This merger consisted of a BH and the first mass gap object (FMG), the nature
of which is unknown (either an NS or a BH). 
\label{tab.events}
\end{table}

\section{Calculations}
\label{calc}

\subsection{Evolutionary calculations}
\label{evol_calculations}

In this paper, we evolve populations of massive star binaries in order to estimate the population of BH-NS binaries, 
focusing on those systems that merge in the Hubble time. All models we present were 
obtained using the {\tt StarTrack} population synthesis code \citep{belczynski2002comprehensive,
belczynski2008compact,belczynski2019evolutionary} under the isolated classical (CE) 
binary evolution scenario. We calculate a series of models with 
different prescriptions for stellar evolution and binary input physics. For each model, 
we use a range of metallicity $Z$ from $0.0001$ up to $0.03$: $32$ sub-models with the 
same input physics but with different metallicity, each calculated for $2 \times 10^{7}$ 
massive binaries. In this section, we review the assumed initial conditions and the 
basic prescriptions used for binary evolution.

We used initial binary parameters from \cite{sana2012binary} as modified by 
\cite{deMink2015}, and stellar winds were adopted from \cite{vink2001mass} and 
\cite{belczynski2010maximum}. We used a $50\%$ binary fraction and we assumed 
a maximum NS mass of $2.5\msun$. We adopted a solar metallicity of $Z_{\odot}=0.014$. 

We adopt three component broken power-law initial mass functions (IMFs) from 
\cite{kroupa2001variation} for the primary (more massive) component in each binary: 
$M_{ZAMS,A} \propto M^{-1.3}$ for $0.08 \leq M<0.5\msun$, 
$\propto M^{-2.2}$ for $0.5 \leq M<1.0\msun$,  and $\propto M^{-\alpha_3}$ for 
$1.0 \leq M<150\msun$, 
where we adopt $\alpha_3=2.3$. The initial mass of the secondary ($M_{ZAMS,B}$) binary 
component is taken from the flat mass ratio distribution in a range $q_0=[0.08/M_{ZAMS,A},1]$,
where $q_0=M_{ZAMS,B}/M_{ZAMS,A}$. The lower range limit is chosen in such a way in 
order to provide second component mass above the hydrogen burning limit ($0.08\msun$).

 In our CE calculations, we use an energy balance approach from 
\cite{webbink1984double}, with updates on binding envelope energy from 
\cite{dominik2012double}. We do not take into account systems in which a CE with a 
Hertzsprung gap donor occurred, as we conservatively assume these merge during the CE phase.  The core-envelope structure of these stars is not 
well known \citep{belczynski2007rarity,ivanova2013common} and the survival 
of these systems during CE is highly uncertain. Additionally, such systems may 
evolve through thermal timescale Roche-lobe overflow (RLOF) rather than through CE 
\citep{Pavlovskii2017}. We adopted an accretion rate  of 
$5\%$ of the Bondi-Hoyle rate~\citep{MacLeod2017} onto the NS and BH during the CE phase.

During stable RLOF with an NS and BH accretor, we  calculate binary evolution and component 
masses using prescriptions from \cite{Mondal2020}. In all the other cases, we 
assume a nonconservative mass transfer, with $50\%$ of the transferred mass being lost 
from a binary with high specific angular momentum $j_{\rm loss}=1$~\citep{podsiadlowski1992presupernova}. 
The remaining $50\%$ of the transferred mass is attached to a companion star. 

Following \cite{fryer2012compact}, we use a neutrino supported convective supernova 
engine to deduce the masses of NSs and BHs in our simulations. We allow for a different 
development time of such an engine. In the rapid supernova model, for intermediate-mass stars, the engine develops 
very quickly ($\sim 100$ ms) and is followed by a supernova explosion and the ejection of stellar 
outer layers , and light NSs form. For massive stars, the 
engine is not able to overcome the weight of in-falling outer stellar layers and stars 
collapse to form rather massive BHs. This naturally creates a mass gap between NSs and BHs
\citep{belczynski2012missing}. In the delayed supernova model, the engine 
develops after a relatively long time ($\sim 500-1000$ ms) after core collapse. 
This allows for significant accretion ($\sim 1-2\msun$) onto light proto-NSs,
and the formation of heavy NSs and light BHs producing a continuous mass spectrum
of compact objects. 

The FMG is typically defined as the range: $\sim 2-5\msun$. 
To reflect the recently discovered (in radio) NSs 
with masses of about $2.1\msun$ (see \cite{cromartie2019very}, \cite{zhang2019implications}), we adopt a slightly narrower range as a definition of the FMG: $2.1-5\msun$ 
in our calculations. 

For massive stars, we take into account pair-instability pulsation supernova  
mass loss. We adopt the weak mass loss that allows for the formation of BHs up 
to $\sim 55\msun$ \citep{belczynski2019evolutionary}. We also allow for the most 
massive stars to be totally disrupted (no BH remnant) by pair-instability supernovae.
 During NS and BH formation, we assume $10\%$ and $1\%$ neutrino mass loss, 
respectively.  

Compact objects during formation may receive natal kicks
\citep{Hobbs2005}.  We use a one dimensional Maxwellian natal kick
magnitude distribution with $\sigma=265\kms$. This generates a three
dimensional average speed of $\sim 420\kms$.  The orientation of the natal
kick is random.  For some models, we allow kicks to be decreased in
magnitude through fallback of matter during compact object
formation. The natal kick decreases with decreasing ejecta mass (increasing 
fallback mass):
\begin{equation}
V_{kick}= (1-f_{fb}) V,
\label{fbkick}
\end{equation}
where $V$ is the kick magnitude drawn from the Maxwellian distribution with
$\sigma=265\kms$, and $f_{fb}$ is a fraction of matter that falls
back onto a compact object after the supernova explosion.  The fallback fraction $f_b$ is obtained from formulae   by
\cite{fryer2012compact}.  It should be noted that this fallback of matter
during core collapse is always used in the calculation of compact
object mass, but is used only for some models in the calculation of natal
kicks.  In other words, when fallback is applied, we assume asymmetric
mass ejection natal kicks, and when fallback is not applied,
we assume asymmetric neutrino emission driven natal kicks. For a more
detailed discussion about natal kicks, see Section~6 of
\cite{belczynski2016compact}.

The natal BH spin magnitude ($a_{\rm spin}=(cJ)/(GM^2)$, where $J$ and $M$ are BH angular momentum 
and mass, respectively) is estimated using one of three different prescriptions of angular momentum 
transport in stellar interiors, depending on the simulation performed. The most efficient angular momentum transport is adopted 
from \cite{fuller2019most,fuller2019slowing} and we assign $a_{\rm spin}=0.01$ for each BH natal spin 
in this scenario. The next scenario employs the standard Tayler-Spruit magnetic dynamo and 
efficient angular momentum transport from \cite{1999A&A...349..189S} as adopted in the 
{\tt MESA} evolutionary code; this model results in $a_{\rm spin}\approx0.05-0.15$ depending on progenitor star mass 
and metallicity. Finally, we also adopt an inefficient angular momentum transport driven by 
meridional currents \citep{ekstrom2012grids} as used in the {\tt Geneva} evolutionary code; 
this results in $a_{\rm spin}\approx0.9$ for low-mass BHs ($\lesssim 20\msun$) and $a_{\rm spin}\approx0.2$ for 
high-mass BHs ($\gtrsim 20\msun$). Full details of the three models are given in 
\cite{belczynski2019evolutionary}.

The natal BH spin magnitude may be affected by tidal interactions
between the stars in progenitor binaries of BH-NS mergers. Therefore,
we allow for tidal spin-up. If a Wolf-Rayet (WR) star forms in a
sufficiently close binary, it is subject to strong tidal
interactions, which affect its rotation in comparison to single
stellar evolution models (see ~\cite{hotokezaka2017implications},
\cite{qin2018spin}). The resultant spin of a BH formed from such a star will
be different from the spin of a BH that was formed either in isolation or
in a wide binary. We adopt natal BH spin magnitudes formed from
tidally affected WR stars from \cite{belczynski2019evolutionary}:
\begin{equation}
a_{\rm spin}=e^{-0.1(P_{\rm orb}/P_0 -1)^{1.1}}+0.125,
\end{equation}
where $P_{\rm orb} [s]$ is the orbital period and $P_0=4000\rm{s}$. We apply this formula 
for $P_{\rm orb}=0.1-1.3$ days. We assume that wider systems (with $P_{orb}>1.3$ days) 
do not experience significant tidal interactions. BHs originating from binaries with 
$P_{orb}<0.1$ days are assigned a spin magnitude of $a_{\rm spin}=1.0$.

Table \ref{modelstab} contains list of our models.  
\begin{table}[h]
\caption{Calculated models, Supernova engine: R/D - rapid/delayed.}
\centering
\begin {tabular}{l|ccr}
\hline\hline
Model & Supernova engine & fallback & BH spin\\
\hline\hline
M230&R&1&MESA\\
M233&R&0&MESA\\
M280&D&1&MESA\\
M283&D&0&MESA\\
M383&D&0&Geneva\\
M483&D&0&Fuller\\
\hline\hline
\end{tabular}
\label{modelstab}
\end{table}

\subsection{BH-NS merger mass ejection and kilonovae}
\label{sec.mejkn}

When an NS is disrupted in a BH-NS merger, the radioactive ejecta can
emit a burst of electromagnetic radiation known as a kilonova.  We do
not consider other potential electromagnetic signals in this study
(e.g., jets producing short gamma-ray bursts considered by
\cite{postnov2019possible}).

During the merger there are two basic scenarios: {\em (i)} the
NS is disrupted outside the BH event horizon (potential EM signal), or
{\em (ii)} the NS is disrupted inside the BH event horizon (no EM
signal). What happens depends on the BH spin, the NS equation of state
($EoS$), the mass ratio of two compact objects and the orientation of the NS
orbit with respect to the BH spin plane. For each BH-NS binary system that
we form in our population synthesis calculations, we consider all these
parameters.  We determine the ejecta mass from \cite{kawaguchi2016models}:
\begin{multline}
\dfrac{M_{ej}}{M_{NS,b}}=\max\{ a_1 Q^{n_1} (1-2C_{NS})C_{NS}^{-1} \\ - a_2 Q^{n_2} \tilde{r}_{ISCO} \left(a_{\rm spin} \cdot cos(i_{\rm tilt}) \right) +  a_3 \left(1-\dfrac{M_{NS}}{M_{NS,b}}\right) +a_4 ,0 \},
\label{Mej_eq}
\end{multline}
where $Q=M_{BH}/M_{NS}$ ($M_{BH}$ is BH mass and $M_{NS}$ is NS mass), 
$i_{\rm tilt}$ is the angle between $BH$ spin and orbital angular momentum, and 
$C_{NS}=GM_{NS}/(c^2R_{NS})$ is the $NS$ compactness parameter that depends 
on the $EoS$. $M_{NS}$ and $R_{NS}$ are neutron star mass and radius, respectively.

\begin{figure}[h]
\includegraphics[width=8.5cm, height=6cm]{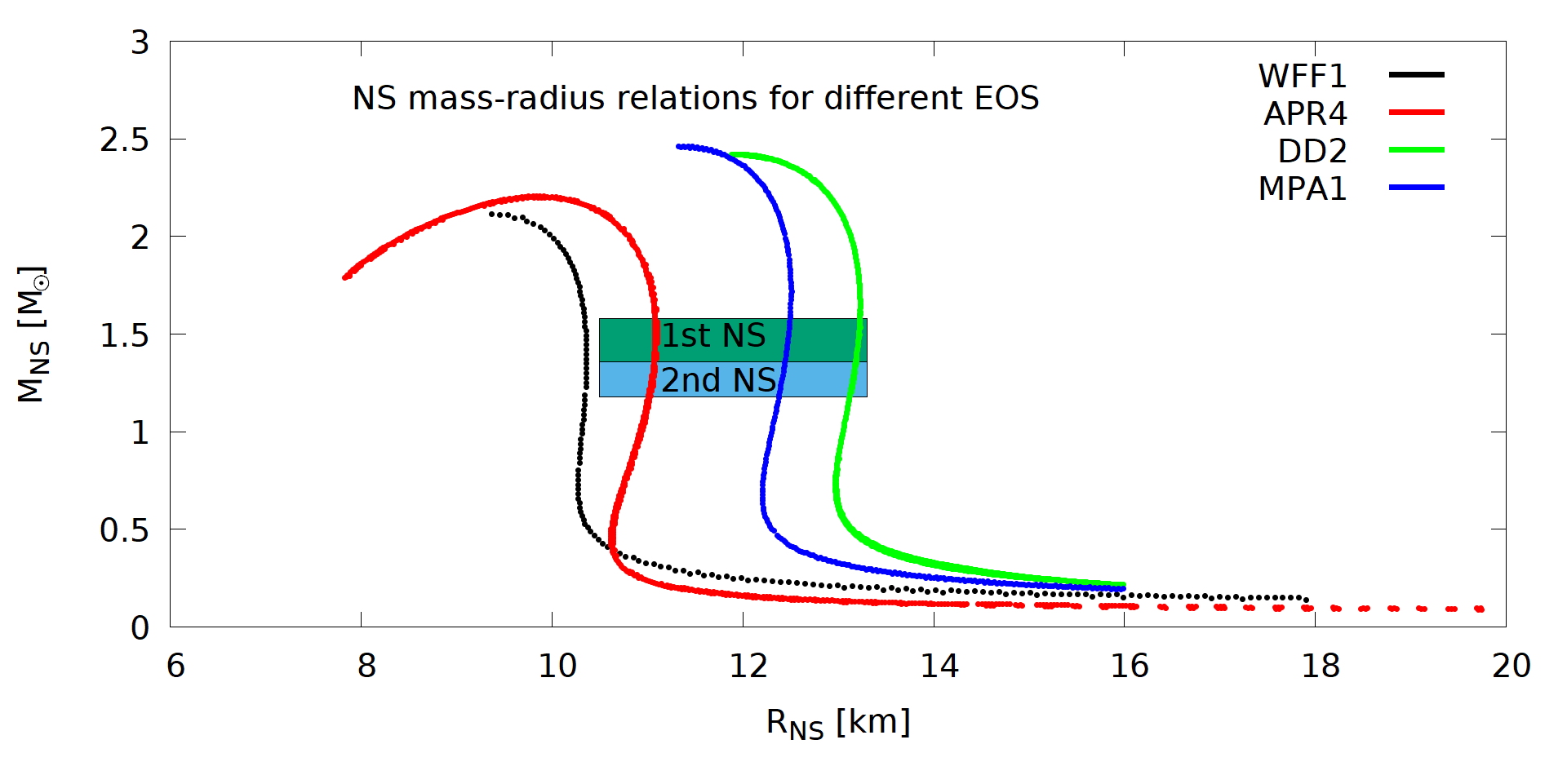}
\caption{
Comparison of the NS mass-radius relation of four different equations of state ($EoSs$)
with the LIGO/Virgo constraints from NS-NS merger GW170817 ($90\%$ credible limits 
shown by green and blue boxes: \cite{abbott2018gw170817}).
We employ the MPA1 $EoS$, which reaches $\sim 2.5 \msun$ (our adopted maximum NS
mass) and agrees with the LIGO/Virgo constraints. Other relations are WFF1 
\citep{wiringa1988equation}, APR4 \citep{akmal1998equation}, and DD2 
\citep{alvarez2016new}. 
}
\label{fig.eos}
\end{figure}

We use the $MPA1$ NS $EoS$, which is consistent with observational constraints on neutron 
star tidal deformation from the first NS-NS merger observation (see Fig.~\ref{fig.eos}), 
as well as neutron star interior composition explorer (NICER) observations \citep{2019ApJ...887L..24M}.  We employ (and this figure shows) mass-radius 
estimates for several  $EoSs$ drawn from
\footnote{\url{https://www3.mpifr-bonn.mpg.de/staff/pfreire/interests.html}}.
The maximum  NS mass for the MPA1 equation of state is $\sim2.46 \msun$. Rather than treat 
objects with masses between $2.46 M_\odot$ and $2.5 M_\odot$ as black holes
($\lesssim 1\%$ of NSs in BH-NS mergers in our models), we assume a constant mass-radius 
relation for objects in this mass range and we treat them as neutron stars. A $2.46\msun$ 
NS has $11.3$ km radius, and then all NSs above this mass (up to $2.5\msun$) have $11.3$ 
\rm{km} radius.

The parameters $a_1$ - $a_4$, $n_1$, and $n_2$ are discussed in
\cite{kawaguchi2016models}.  $M_{NS,b}$ is the total baryonic mass of
an NS.  We use the relation between $M_{NS}$ and $M_{NS,b}$ for
nonrotating $NS$ from \cite{gao2020relation} and
\cite{gupta2019black}, lowered by $1.8\%$ to account for  the relative error of NS
 baryonic mass estimation (a lower $M_{NS,b}$ provides more realistic ejecta masses).
$\tilde{r}_{ISCO}$ is the innermost stable
circular orbit radius normalized by
$M_{BH}$ (e.g., from \cite{foucart2018remnant}): 
\begin{equation}
\tilde{r}_{ISCO}=\dfrac{r_{ISCO}}{GM_{BH} c^{-2}}=3+Z_2-{\rm sign}(a_{\rm spin})\sqrt{(3-Z_1)(3+Z_1+2Z_2)},
\end{equation}
where $Z_1$ and $Z_2$ are functions of $BH$ spin, and $Z_1=1+(1-a_{\rm spin}^2)^{1/3}((1+a_{\rm spin})^{1/3} +(1+a_{\rm spin})^{1/3})$ 
and $Z_2=\sqrt{3a_{\rm spin}^2+Z_1^2}$. However, we use the projection of BH spin on the binary orbital momentum direction to calculate 
$\tilde{r}_{ISCO}$ ($\tilde{r}_{ISCO}(a_{\rm spin} \cdot \cos(i_{\rm tilt}))$).

Kilonova emission remains an active area of research and the peak
luminosity depends on a wide range of factors concerning the properties of the
ejecta, such as their composition, their morphology, and the viewing angle (see
\cite{wollaeger2019impact}, \cite{wollaeger2018impact},
\cite{zhu2018californium}, \cite{fontes2020line}, and \cite{korobkin20}).
For this paper, we use the following parameterized formula for the o-band ($1260-1360$\,nm)
peak luminosity:
\begin{equation}
L_{peak}=f \times 10^{40} \left( \dfrac{M_{ej}}{0.01\msun}\right)^{0.2} \left( \dfrac{v}{0.1 c}\right)^{0.2}
\rm{erg\ s^{-1}},
\label{lpeak}
\end{equation}
where $f$ is varied to match the wide range of current results: $0.3\leq f \leq4.5$. We assume that $f$ 
has uniform distribution in this range. For ejecta velocity, we use the formula from
\cite{kawaguchi2016models}:
\begin{equation}
v/c=0.01533 Q + 0.1907.
\label{vej}
\end{equation}

We calculate apparent flux from the kilonova for each merger event with mass ejection. We assume 
that mass ejection is isotropic. Therefore, the flux can be calculated as:
\begin{equation}
F=\dfrac{L_{peak}/(1+z)^2}{4 \pi D_L^2},
\label{fluxx}
\end{equation}
where $z$ is redshift and $D_L$ is the luminosity distance of a given merger. 

This flux corresponds to the apparent o-band magnitude on Earth. The apparent kilonova magnitude can 
be calculated using Pogson's equation:
\begin{equation}
M_{kn}=m_{\rm Vega}-2.5 \log_{10} \big( \frac{F}{F_{Vega}} \big),
\label{eq.magg}
\end{equation}
where $m_{\rm Vega}=0.026 \rm{mag}$ is the apparent Vega magnitude as seen from Earth, 
$F_{Vega}=2.18072\times 10^{-8}\ \rm{W\ m^{-2}}$ is flux from Vega, and $F$ is flux from the 
kilonova.

\subsection{Cosmology}
\label{sec.cosmo}

We assume a flat Universe with $\Omega_M=0.3$, $\Omega_{\Lambda}=0.7$, $\Omega_k=0$ 
(\cite{ade2016planck}), and Hubble constant $H_0=68.7\ \rm{km\ s^{-1}\ Mpc^{-1}}$. 
This gives the age of Universe $t_0 \sim 13.7 \rm{Gyr}$ \citep{aghanim2018planck,hinshaw2013nine}. 

We assume that binary stars (Population I and II) are formed in the redshift range 
$z=0-15$. The star formation rate density history is adopted from \cite{madau2017radiation}:
\begin{equation}
{\rm sfrd}(z)=0.015 \frac{(z+1)^{2.6}}{1+((z+1)/3.2)^{6.2}}\ \msun\ {\rm Mpc}^{-3}\ {\rm yr}^{-1}.
\end{equation}

We adopt an average cosmic metallicity evolution from \cite{madau2017radiation} and 
we adopt a Gaussian distribution (with $0.5$ dex sigma) of metallicity
around the average at each redshift \citep{belczynski2019evolutionary}.
After evolution, some binaries form BH-NS systems and some of these systems merge. 
We record their merger redshifts (and corresponding distances). We also calculate 
an intrinsic BH-NS merger rate density and an intrinsic merger rate as a function of 
redshift for all of our models, following the method described in detail by 
\cite{belczynski2016compact}.

\subsection{Result presentation}
\label{presentation}

The results of our simulations are presented in the context of intrinsic properties of binary 
systems, as well as the properties of the subset of systems likely to be detected by 
LIGO/Virgo. We mostly focus on low redshift systems ($z<1$) because BH-NS mergers will not 
be detected by LIGO/Virgo above redshift $z=1$ even at full design sensitivity. We also 
present results based on the current sensitivity limits of the LIGO/Virgo O3 science run.

We present the intrinsic fraction of BH-NS mergers that have compact object or 
objects in the FMG, which we estimate  from:
\begin{equation}
\eta_1= \frac{Rd_{fmg,z<1}}{Rd_{z<1}},
\label{BHNS_fit1}
\end{equation}
where $Rd_{fmg,z<1}$ ($\gpy$) is the intrinsic merger rate density for
BH-NS mergers that occur for redshifts $z<1$ and that have two, one,
or no objects within the FMG (assumed $2.1-5\msun$ range,
see Sec.~\ref{evol_calculations}).  If there is only one compact
object within the FMG, we identify whether it is an NS or a
BH (using our $2.5\msun$ maximum mass limit for an NS). $Rd_{z<1}$
($\gpy$) is the intrinsic merger rate density for all $z<1$ BH-NS mergers.

To present results for mass ejection in mergers of BH-NS binaries, we use a fraction 
of intrinsic merger rate density. The intrinsic fraction of $z<1$ BH-NS mergers 
with some mass ejection is found from: 
\begin{equation}
\eta_2= \frac{Rd_{Mej,z<1}}{Rd_{z<1}},
\label{BHNS_fit2}
\end{equation}
where $Rd_{Mej,z<1}$ ($\gpy$) is the intrinsic merger rate density for BH-NS mergers 
that occur at redshifts $z<1$ and has an ejecta mass $M_{ej}>0.001\msun$ or 
$M_{ej}>0.01\msun$, and $Rd_{z<1}$ ($\gpy$) is the intrinsic merger rate density for 
all $z<1$ BH-NS mergers. 

Using simple mass scaling, we assess the detectability of BH-NS mergers in gravitational waves by LIGO/Virgo.  We assume that BH-NS mergers are only detectable  if the signal-to-noise 
ratio (S/R) in one LIGO/Virgo detector is larger than $8$, and we estimate a fiducial S/R from:
\begin{equation}
S/R=8.0\dfrac{120\rm{Mpc}}{D_L} \left(\frac{M_{chirp}}{1.2\msun}\right)^{5/6},
\label{SNR}
\end{equation}
where $M_{chirp}=(m_1 m_2)^{3/5} (m_1 + m_2)^{-1/5}$ is the chirp mass of the BH-NS system, 
and $m_1$ and $m_2$ are BH-NS components masses. In this formula, we assume that the 
current O3 LIGO/Virgo detection range for an NS-NS merger with a typical chirp mass 
of $1.2\msun$ is $120$ Mpc \citep{abbott2018prospects}), and we average over the source orientation and sky location. This gives us a detection 
rate of BH-NS mergers by LIGO/Virgo: $R_{ligo}$ yr$^{-1}$.
The fraction of BH-NS mergers that have compact objects in the FMG that are detectable by LIGO/Virgo is calculated from:
\begin{equation}
\eta_3= \frac{R_{fmg,ligo}}{R_{ligo}},
\label{BHNS_fit3}
\end{equation}
where $R_{fmg,ligo}$ (yr$^{-1}$) is the LIGO/Virgo detection rate of BH-NS mergers
that have two, one or no objects within the FMG. 

We also calculate the fraction of LIGO/Virgo detectable BH-NS mergers ($S/R>8$) that show some mass ejection
($M_{\rm ej}$) during the merger process as: 
\begin{equation}
\eta_4= \frac{R_{Mej,ligo}}{R_{ligo}},
\label{BHNS_fit4}
\end{equation}
where $R_{ligo}$ (yr$^{-1}$) is the LIGO/Virgo detection rate of BH-NS mergers, and 
$R_{Mej,ligo}$ (yr$^{-1}$) is the detection rate of BH-NS mergers with $M_{\rm ej}$
larger than a specified value (i.e., $M_{\rm ej}>0.001\msun$).

BH-NS mergers that are detectable as kilonovae are defined as follows: first, they need 
to be detectable by LIGO/Virgo ($S/R>8$) and then they need to have enough 
mass ejecta escaping with appropriate speed during the merger to produce emission
bright enough to be detected by a given telescope (see eq. \ref{lpeak}).
For each BH-NS merger with mass ejection (and kilonova), we use 
its apparent magnitude ($M_{\rm kn}$, see eq.~\ref{eq.magg}) and compare it with the 
threshold sensitivity of three different telescopes that are employed in the search for 
kilonovae.

We consider a range of telescopes, from small instruments that can cover a large fraction 
of the sky, such as ATLAS (2 telescopes, each with a diameter of $0.5$m) with a maximum reach of 
$M_{\rm max}=19.5$mag (exposure time $0.5$min), to the medium-size Canada-France-Hawaii 
Telescope (CFHT) (1 telescope with a diameter of $3.6$m) with $24.1$mag limit (exposure 
time $60$min), and the large Subaru telescope (1 telescope with a diameter of $8.2$m) with a limit of 
$26.0$mag (exposure time $4-8$min).
These exposure times are adopted from the official cameras’ specifications as characteristic times at certain bands.

 ATLAS's detection limit is given 
directly for o-band, while the CFHT and Subaru limits are adopted for J-band ($WIRCam$ camera 
limit, band centered at $1220$ nm) and $Ic$ filter ($Suprime-Cam$ camera limit (see 
\cite{miyazaki2002subaru}), filter centered at $806$ nm), respectively. If a kilonova is 
brighter than the threshold for a given telescope:
\begin{equation}
M_{\rm kn}<M_{\rm max}
\label{kill} 
\end{equation}
then we call it a detectable EM/kilonova counterpart to a given LIGO/Virgo BH-NS 
merger signal. Then we calculate the fraction of detectable BH-NS mergers as 
kilonova that are brighter than chosen telescope brightness threshold (see 
eq.~\ref{kill}) in all LIGO/Virgo detectable BH-NS mergers from:
\begin{equation}
\eta_5= \frac{R_{kn,ligo}}{R_{ligo}},
\label{BHNS_fit5}
\end{equation}
where $R_{kn,ligo}$ (yr$^{-1}$) is the LIGO/Virgo detection rate of BH-NS mergers
with detectable kilonovas, and $R_{ligo}$ (yr$^{-1}$) is the LIGO/Virgo detection 
rate of BH-NS mergers. 

It should be noted that the above scheme does not account for any localization issues that 
may arise during the search for kilonovae associated with LIGO/Virgo sources (e.g., 
\cite{Nissanke2013,Gomez2019}). This approximation serves only as a guide to inform 
us whether any given kilonova is bright enough to be detected (with some typical 
exposure time) by a given telescope, if this telescope was pointed right at the 
kilonova. Additionally, we use peak brightness, so this is a very optimistic 
approximation of kilonova detectability.

\section{First mass gap objects in BH-NS mergers}
\label{sec.gap}

\begin{figure}[h]
\includegraphics[width=8.5cm, height=6cm]{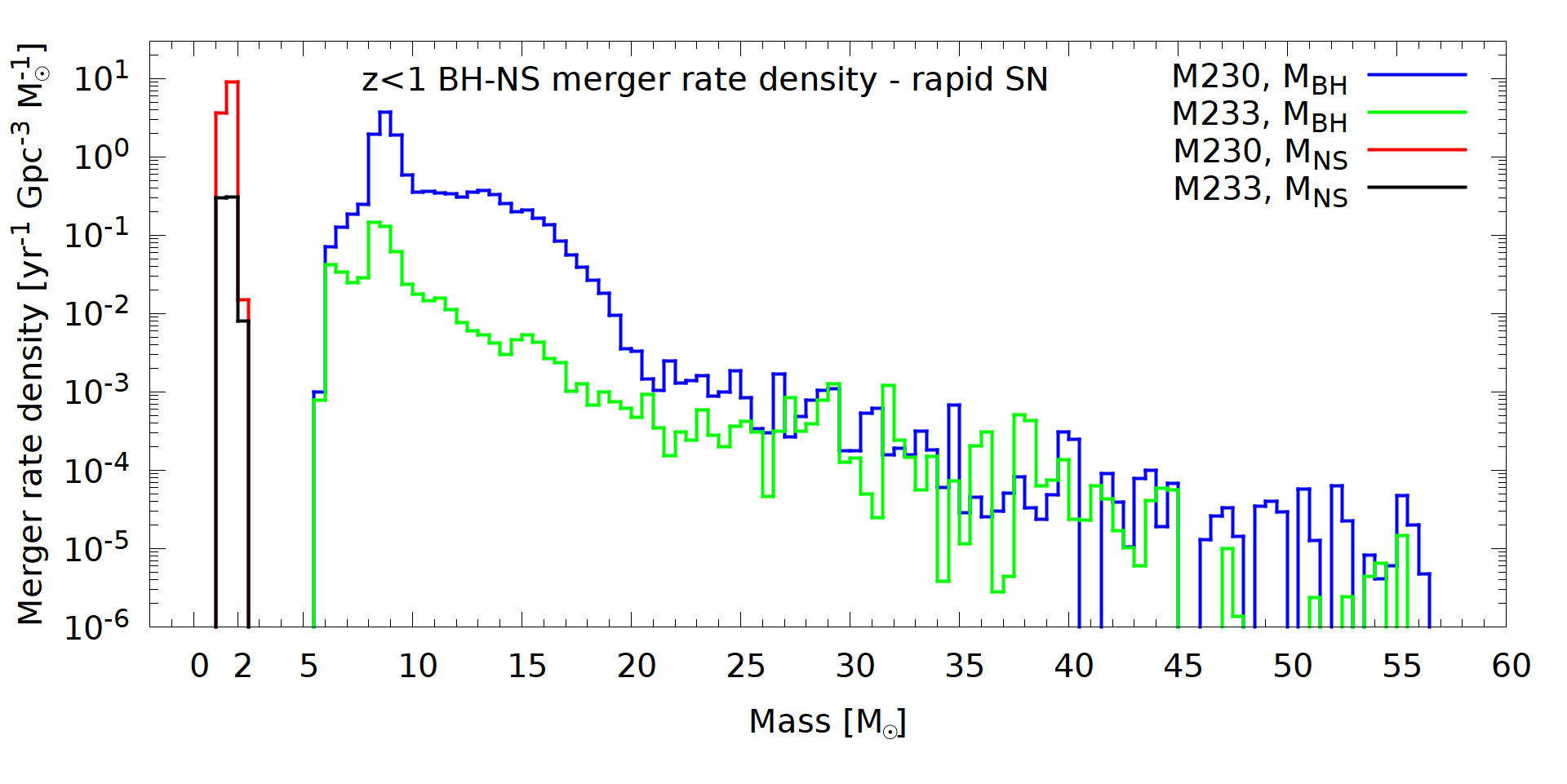}
\caption{Mass distribution of compact objects in BH-NS mergers in the low-redshift Universe 
($z<1$) for models that employ a rapid supernova engine (M230, M233). There is a very 
clear mass gap between NSs and BHs.
It should be noted that our data become noisy for masses
above $\gtrsim 20\msun$. However, apparent decline in merger rate density for
these higher masses is real and expected in our models.}
\label{hist_rapid}
\end{figure}

\begin{figure}[h]
\includegraphics[width=8.5cm, height=6cm]{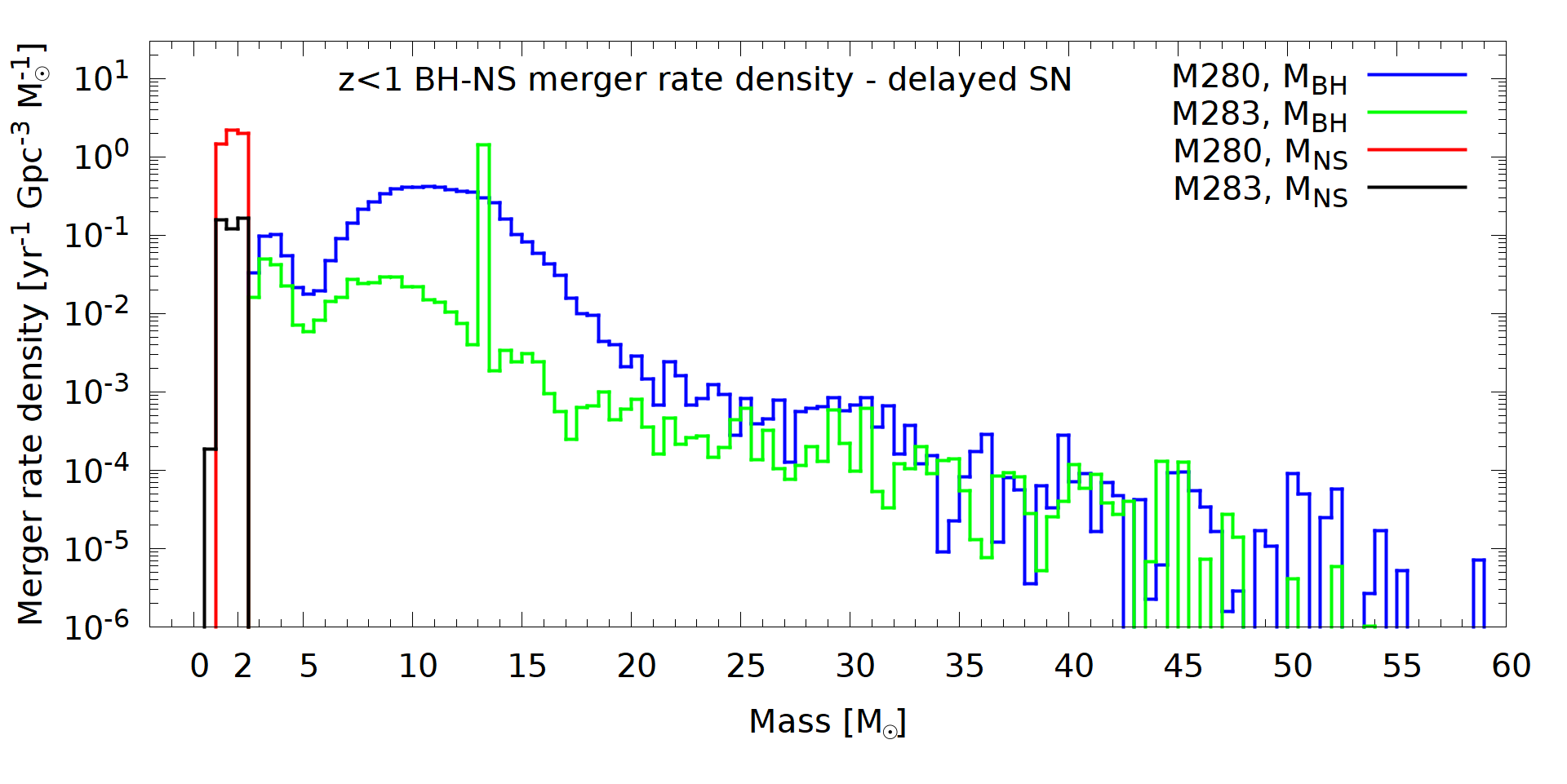}
\caption{Mass distribution of compact objects in BH-NS mergers in the low-redshift Universe 
($z<1$) for models that employ a delayed supernova engine (M280, M283). There is no mass 
gap between NSs and BHs. It should be noted that our data become noisy for masses
above $\gtrsim 20\msun$, although apparent decline in merger rate density for
these higher masses is real and expected in our models.}
\label{hist_delayed}
\end{figure}

First we study the  BH-NS merger compact object mass distribution ($z<1$) for
several evolutionary models (M230, M233, M280, and M283) using both 
rapid (Figure~\ref{hist_rapid}) and delayed
(Figure~\ref{hist_delayed}) supernova engine prescriptions for remnant
masses. As expected, the FMG is clearly visible in models
that employ a rapid supernova engine (e.g., M230 and M233). Therefore, for these
models, we do not expect BH-NS mergers with compact objects of comparable
mass. However, for models that employ a delayed supernova engine (e.g., M280 and M283),
there is no mass gap between NSs and BHs, and we predict some BH-NS mergers
with comparable mass compact objects. This is important in context of mass 
ejection in BH-NS mergers (see Sec.~\ref{sec.em}).

In Figures \ref{hist_rapid} and \ref{hist_delayed} we also clearly see the 
effect of natal kicks on the merger rate density of BH-NS systems. For models with 
decreased natal kicks (e.g., M230 and M280) the rates are higher by about an order of 
magnitude than for models with high natal kicks (e.g., M233 and M283). This is a 
consequence of BH-NS progenitor binaries being disrupted more easily in models in 
which we employed higher natal kicks. One can also see some noisy features
that appear at high mass ends. Those are the results of low statistics.

In Table~\ref{MG_intrinsic} we list fractions of BH-NS mergers that
have both compact objects within the FMG ($FMG_{\rm both}$),
only a BH in the gap ($FMG_{\rm BH}$), only an NS in the gap ($FMG_{\rm NS}$), 
or no components in the gap ($FMG_{\rm none}$). This table
contains BH-NS mergers that take place for redshifts with $z<1$ and the
fractions are calculated using merger rate densities (see
eq.~\ref{BHNS_fit1}). For models that the employ rapid supernova engine, there are
no BH-NS mergers with objects in the FMG (assumed
$2.1-5\msun$ range). For models that employ the delayed supernova engine
remnant prescription, about half or a little more than half ($0.576-0.708$) of
BH-NS mergers do not have any mass gap objects.  With these models,
those that employ high natal kicks independent of compact object mass
(M283, M383, and M483) show a significant fraction of BH-NS mergers with
both compact objects within the mass gap ($0.197-0.212$). Since all
compact objects have an equal probability of getting a high natal kick and
disrupting the progenitor binary, the mass function of these compact objects
falls steeply off with mass (approximately following the initial mass
function of the stars); see the  black (NS) and green (BH) lines in
Figure~\ref{hist_delayed}. That results in a significant fraction of
BH-NS mergers with both compact objects within the FMG. In
contrast, for our model in which natal kicks decrease with compact object
mass (M280), this fraction is very small ($0.036$) because  
many BH-NS systems with heavy ($\sim 10\msun$) BHs outside the FMG
are not disrupted by the kick; see the blue line in
Figure~\ref{hist_delayed}.  Depending on the model, significant fractions
of BH-NS mergers ($0.211-0.256$) have one compact object (whether it
is an NS or a BH) within the FMG.  Although we differentiate
between NSs and BHs in the mass gap in Table~\ref{MG_intrinsic}, we
note that this division is arbitrary as we simply assume that any
compact object above $2.5\msun$ is a BH.

In Table~\ref{MG_ligo} we list fractions of BH-NS mergers with compact 
objects within the FMG, but only for systems that are detectable 
by LIGO/Virgo (S/R>8; see eq.~\ref{BHNS_fit3}). The detectable LIGO/Virgo 
population is predicted to have a significant fraction of mergers ($0.199-0.334$) with 
at least one compact object within the mass gap for delayed supernova engine models (M280, 
M283, M383, and M483). A small but noticeable fraction ($0.025-0.186$) of BH-NS 
mergers are found to have both compact objects within the gap. Compared with
the intrinsic population (see Table~\ref{MG_intrinsic}), the LIGO/Virgo 
detectable population has a smaller fraction of BHs and a larger fraction of NSs 
in the gap. This simply reflects the fact that LIGO/Virgo can detect heavier
objects from larger distances and that light BHs within the gap tend to be less
represented in the LIGO/Virgo detectable population, while NSs need to be 
heavy to be within the gap. Thus, NSs stand out more in this population. 

The fraction of FMG+FMG mergers among the entire LIGO/Virgo detectable BH-NS merger 
population is significantly smaller for the low BH natal kick model (M280: $\sim 0.04$) 
than for high kick models (M283, M383, and M483: $\sim 0.2$; see Table~\ref{MG_ligo}). FMG+FMG 
systems tend to survive high natal kicks more often than other BH-NS merger progenitors 
because of their specific evolutionary history that typically leads to small separations 
before supernova explosions (see Sec.~\ref{3.5+2.1}). Therefore, models with high natal 
kicks contain higher fractions of FMG+FMG systems, which are mergers that are
typically accompanied by mass ejection, (see Sec.~\ref{sec.mj}) and thus show 
kilonova emission (see Sec.~\ref{sec.em}). However, these models have small BH-NS merger 
rates (see Sec.~\ref{sec.rate}).

We find that BH-NS mergers may still be detectable at distances as far as $z\sim 0.1$ 
with LIGO/Virgo during O3 ($S/R>8$).  Due to GW selection effects, the  most distant 
mergers would be the most massive ones, for example: $2.2\msun$ NS + $30.8\msun$ BH.  

\begin{table}[h]
\caption{Intrinsic fraction of ($z<1$) BH-NS mergers $\eta_1$ in which both
($\rm{FMG_{both}}$) components, 
only a BH ($\rm{FMG_{BH}}$) component, only an NS ($\rm{FMG_{NS}}$) component, or no ($\rm{FMG_{none}}$) 
components are within the FMG.}
\centering
\begin {tabular}{l|ccccr}
\hline\hline
Model&$ \rm{FMG_{both}}$ & $\rm{FMG_{BH}}$ & $\rm{FMG_{NS}}$ & $\rm{FMG_{none}}$ \\
\hline\hline
M230 & 0 & 0 & 0 & 1\\
M233 & 0 & 0 & 0 & 1\\
M280 & 0.036 & 0.019 & 0.237 & 0.708\\
M283 & 0.197 & 0.110 & 0.101 & 0.592\\
M383 & 0.200 & 0.098 & 0.119 & 0.582\\
M483 & 0.212 & 0.107 & 0.104 & 0.576\\
\hline\hline
\multicolumn{5}{l}{Note: see eq.~\ref{BHNS_fit1}. The same results expressed in terms of merger}\\
\multicolumn{5}{l}{rates can be found in the  appendix; see  Table~\ref{rate.MG_intrinsic}.}
\end{tabular}
\label{MG_intrinsic}
\end{table}

\begin{table}[h]
\caption{Fraction of LIGO/Virgo ($S/R>8$) BH-NS systems $\eta_3$ where both ($\rm{FMG_{both}}$) components, 
only a BH ($\rm{FMG_{BH}}$) component, only an NS ($\rm{FMG_{NS}}$) component, or no ($\rm{FMG_{none}}$) 
components are within the FMG.}
\centering
\begin {tabular}{l|ccccr}
\hline\hline
Model& $\rm{FMG_{both}}$ & $\rm{FMG_{BH}}$ & $\rm{FMG_{NS}}$ & $\rm{FMG_{none}}$ \\
\hline\hline
M230 & 0 & 0 & 0 & 1\\
M233 & 0 & 0 & 0 & 1\\
M280 & 0.025 & 0.002 & 0.332 & 0.641\\
M283 & 0.186 & 0.011 & 0.188 & 0.614\\
M383 & 0.171 & 0.024 & 0.226 & 0.578\\
M483 & 0.159 & 0.010 & 0.193 & 0.638\\
\hline\hline
\multicolumn{5}{l}{Note: see eq.~\ref{BHNS_fit3}.}\\
\end{tabular}
\label{MG_ligo}
\end{table}

\section{Mass ratio of BH-NS systems}
\label{q}

\begin{figure}[h]
\includegraphics[width=8.5cm, height=6cm]{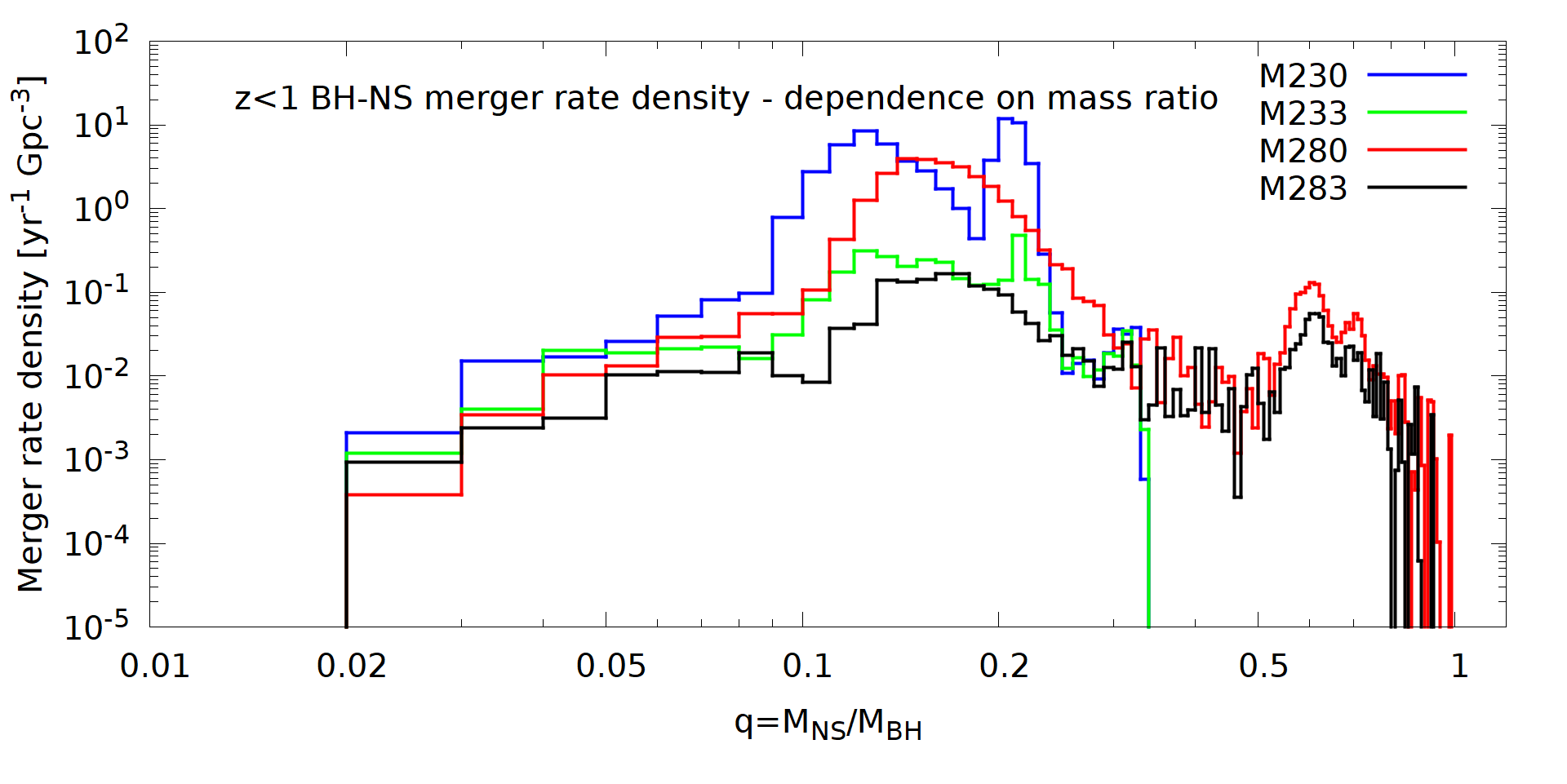}
\caption{Mass ratio distribution for BH-NS mergers in the low-redshift Universe ($z<1$).
For each $BH-NS$ system, we calculated the mass ratio
$q=M_{NS}/M_{BH} \in[0,1]$.
Values indicate merger rate densities estimated from the systems
within each q bin, that is to say $dRate$ $dt^{-1}dV^{-1}dq^{-1} \Delta q$, with $\Delta q=0.01$.
For transparency purposes, we use simple merger rate density as the vertical axis label.
Most mergers have rather small mass ratios $q \sim 0.1-0.2$, but there are 
also extreme mass ratio systems such as $56\msun$ BH + $1.3\msun$ NS ($q=0.023$).
For models with the delayed supernova engine (M280 and M283), there is a secondary peak 
for high mass ratios $q>0.4$, which consists mostly of double compact objects with both components 
in the FMG (see Sec.~\ref{q}).} 
\label{fig.q}
\end{figure}

In Figure~\ref{fig.q} we show the mass ratio distribution (mass of an NS to a BH: 
$q=M_{NS}/M_{BH}$) for models with a rapid supernova engine (M230 and M233) and models 
with delayed supernova engine (M280 and M283). This figure includes the population of BH-NS 
mergers at redshifts $z<1$ and shows the intrinsic mass ratio expressed by the merger 
rate density in bins of $\Delta q=0.01$.

As seen immediately from Figure~\ref{fig.q}, the majority of BH-NS mergers have
small mass ratios $q \sim 0.1-0.2$ independent of model assumptions. This 
comes from the fact that NSs have typical masses of $1-2\msun$ and the majority 
of BHs found in BH-NS mergers in our models have a typical mass of $\sim 10\msun$ 
(see Figs.~\ref{hist_rapid} and ~\ref{hist_delayed}). Since the rapid supernova engine 
does not populate the mass gap, when adopting this model, we thus predict 
no mergers with a mass ratio larger than $q \sim 0.4$. However, for models with 
a delayed supernova engine, the mass ratio distribution shows a secondary peak for $q \sim 0.5-0.8$. 
These are BH-NS mergers with both compact objects within the FMG, as such 
systems must have $q \gtrsim 0.4$.  

\begin{figure}[h]
\includegraphics[width=9cm, height=6cm]{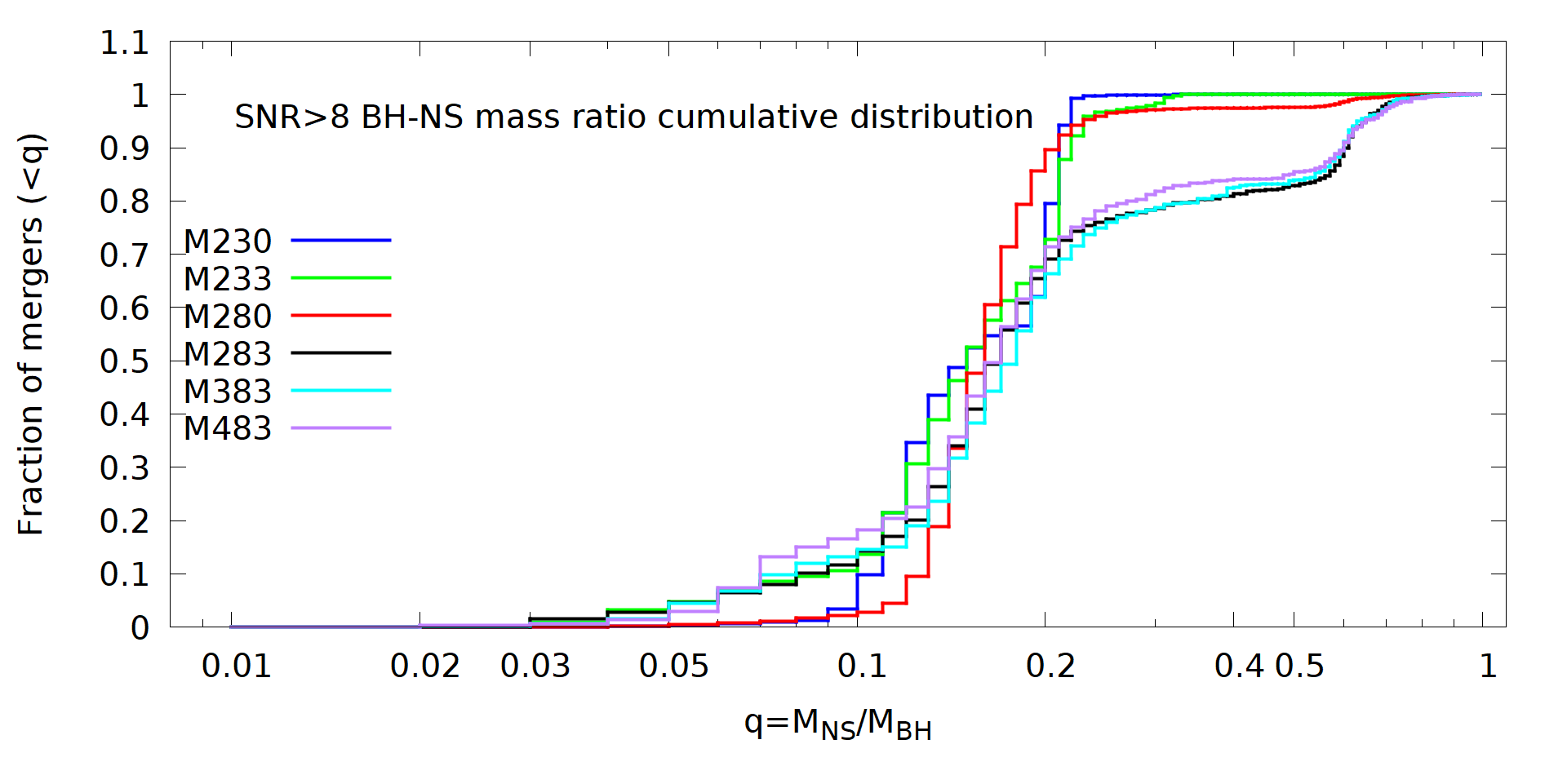}
\caption{Cumulative distribution of LIGO/Virgo detectable ($S/R>8$) BH-NS 
mergers in terms of mass ratio (see also Table~\ref{tab.extreme}).}
\label{distr_q_det}
\end{figure}

In Figure~\ref{distr_q_det} we show cumulative mass ratio distribution for LIGO/Virgo 
detectable ($S/R>8$) BH-NS mergers. In Table~\ref{tab.extreme} we list fractions of 
LIGO/Virgo detectable BH-NS mergers ($S/R>8$) that have a mass ratio smaller than $1/50$, 
$1/30$, $1/20$,  and $1/10$ for all our models. Additional models that include BH-NS mergers 
with input physics assumptions that are different to those used in this study may be found at 
\footnote{\url{www.syntheticuniverse.org}} under the tab “Download/2020: Double Compact 
Objects/Belczynski et al. 2020". 

Our results show that it is possible to create extreme mass ratio systems 
$q<0.05$. For example, the most extreme systems detectable by LIGO/Virgo ($S/R>8$) 
in our models are: 
M230 $q=0.021$ ($M_{BH}=55.0\msun$, $M_{NS}=1.2\msun$), 
M233 $q=0.023$ ($M_{BH}=49.0\msun$, $M_{NS}=1.1\msun$),
M280 $q=0.025$ ($M_{BH}=52.5\msun$, $M_{NS}=1.3\msun$),
M283 $q=0.029$ ($M_{BH}=29.9\msun$, $M_{NS}=0.9\msun$), 
M383 $q=0.022$ ($M_{BH}=59.5\msun$, $M_{NS}=1.3\msun$),
and M483 $q=0.021$ ($M_{BH}=60.1\msun$, $M_{NS}=1.3\msun$). 

In the next subsection we discuss in detail the formation of one extreme mass 
ratio system ($38.9\msun + 1.3\msun$; Sec.~\ref{50+1}), as infrequent as it may 
be. We follow with a detailed description of the formation of a BH-NS merger with  a
typical mass ratio ($14.7\msun + 1.8\msun$; Sec.~\ref{14.7+1.8}), and finish 
with a description of a BH-NS merger with comparable mass components, both within 
the FMG ($3.5\msun + 2.1\msun$; Sec.~\ref{3.5+2.1}).

\begin{table}[h]
\caption{Fraction of LIGO/Virgo detectable ($S/R>8$) BH-NS mergers with extremely 
small ratios and extremely small mass ratios.}
\centering
\begin {tabular}{l|cccc}
\hline\hline
Model & q<1/50 & q<1/30 & q<1/20 & q<1/10 \\
\hline\hline
M230 & 0 & 0.0003 & 0.0017 & 0.0332\\
M233 & 0 & 0.0028 & 0.0319 & 0.1071\\
M280 & 0 & $<10^{-4}$ & 0.0022 & 0.0208\\
M283 & 0 & 0.0017 & 0.0274 & 0.1171\\
M383 & 0 & 0.0023 & 0.0149 & 0.1321\\
M483 & 0 & 0.0047 & 0.0139 & 0.1650\\
\hline\hline
\multicolumn{5}{l}{Note: The same results expressed in terms of merger rates}\\
\multicolumn{5}{l}{can be found in the appendix; see Table~\ref{rate.extreme}.}\\
\end{tabular}
\label{tab.extreme}
\end{table}

\subsection{Extreme mass ratio BH-NS: $38.9+1.3\msun$ merger}
\label{50+1}

\begin{figure}[h]
\includegraphics[width=8.5cm, height=15cm]{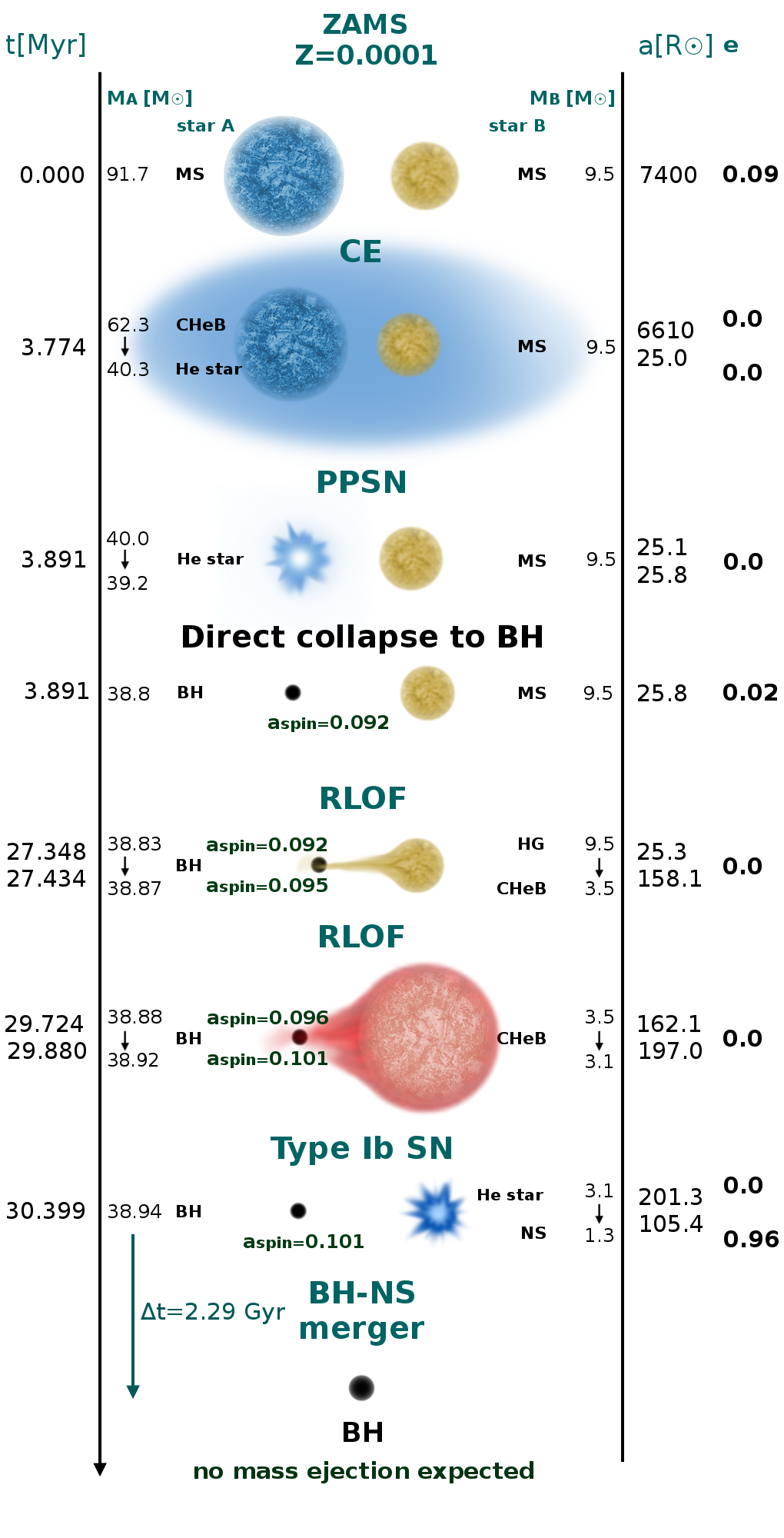}
\caption{
Evolutionary route leading to the formation of the $38.9\msun + 1.3\msun$ 
($q=0.033$) BH-NS merger for metallicity $Z=0.0001$ (model M280). The evolution 
from ZAMS to the formation of the BH-NS system takes $\sim 30.4$ Myr, and then it 
takes $\sim 2.29$ Gyr for the system to in-spiral (angular momentum loss due to 
the emission of gravitational waves), leading to the final merger of the two compact objects. 
(CHeB: core helium burning star, He star: naked helium star, HG: Hertzsprung 
gap star). No tidal spin-up is expected due to large orbital separations when
binary components are naked helium stars: $a\sim 25\rsun$ ($P_{\rm orb}=2.1$d; 
star A) and $a\sim 200\rsun$ ($P_{\rm orb}=60$d; star B) (see Sec. 
\ref{evol_calculations}). 
}
\label{fig.evol1}
\end{figure}

In this study we use the terms primary and secondary component for binary components 
with higher and lower zero age main sequence (ZAMS) mass, respectively.
Our calculations show that BH-NS mergers (independent of their mass ratio)
follow a similar evolutionary history. In Figure~\ref{fig.evol1} we show the major
stages of evolution of a massive binary (primary $M_1=91.7\msun$, secondary 
$M_2=9.5\msun$) that ultimately produces an extreme mass ratio BH-NS merger 
($M_{BH}=38.9\msun$, $M_{NS}=1.3\msun$: $q=0.033$).
Below we describe the major evolutionary stages.

The evolution starts with a very massive primary ($M_{\rm A}=91.7\msun$)
and an intermediate mass secondary ($M_{\rm B}=9.5\msun$) on a very wide orbit 
(semimajor axis $a=7400\rsun$) with a small eccentricity ($e=0.09$)
Next, the primary becomes a core helium burning star while the secondary is 
still on the main sequence.
Then, a CE phase is initiated by expansion of the primary; the primary
      becomes a naked helium star after H-rich envelope ejection.
Afterwards, the primary experiences a pair-instability pulsation supernova mass loss 
      and immediately collapses directly to a BH.
The secondary becomes a Hertzsprung Gap (HG) star and, $14$ kyr later, it 
      initiates a stable Roche lobe overflow mass transfer that ends when 
      the secondary is in the core helium burning stage, making a blue loop on the H-R diagram 
      (radius decreases).
Then, the Roche-lobe overflow restarts while the secondary is still in the core helium burning
stage  and it is moving redwards on the H-R diagram loop (radius increases).
The Roche lobe overflow ends when the secondary loses most of its H-rich envelope.
Afterwards,
 the secondary becomes a naked helium star and explodes in a Type Ib (core-collapse) 
      supernova, forming an NS.

A highly eccentric BH-NS binary with an extreme mass ratio is formed.
After more than 2 billion years of tightening the orbit,
the merger of the BH with the NS leads to a burst of gravitational waves, but
      there is no mass ejection during the merger (no kilonova).

There is nothing specifically unusual in the formation of extreme mass ratio 
BH-NS mergers. Similar evolutionary routes are responsible for nonextreme 
BH-NS mass ratio systems (see \cite{klencki2018impact}). BH-NS progenitors 
experience two major binary interactions (CE and RLOF) and are subject to one 
strong supernova explosion forming an NS. The natal kick ($243\kms$ in the case of 
the example shown in Fig.~\ref{fig.evol1}) produced during the supernova can make the BH-NS 
system highly eccentric. This reduces the time to final merger,  allowing
such systems to merge within Hubble time and to be detected by LIGO/Virgo. 
However, for extreme mass ratio systems there is no mass ejection during the merger 
and therefore these mergers are not expected to produce EM (e.g., kilonova
or short GRB) signals (see Sec.~\ref{sec.em}).

\subsection{Typical mass ratio BH-NS: $14.7+1.8\msun$ merger}
\label{14.7+1.8}

There are several variations to the formation of BH-NS mergers in respect to the
evolution presented in Figure~\ref{fig.evol1}. Here we pick one such variation 
to show the formation of a typical mass ratio ($q=0.12$) BH-NS merger in our 
simulations. 

The formation of a $14.7+1.8\msun$ BH-NS merger at metallicity $Z=0.0005$ (model
M230) involves the evolutionary stages, which we describe below.

The evolution starts with a massive primary ($M_{\rm A}=35.8\msun$)
 and an intermediate mass secondary ($M_{\rm B}=20.6\msun$) on a wide 
($a=6641\rsun$) and eccentric orbit ($e=0.56$).
At some point, the primary becomes an asymptotic giant branch (AGB) star, while the secondary is 
still on the main sequence. As a result, the CE phase is initiated by the primary,
which loses about half of its mass.
Then, the primary undergoes a direct collapse (only a $0.15\msun$ mass loss in neutrinos, 
 no supernova explosion) to a BH and we assume no natal kick. Next, the secondary becomes
a core helium burning star, initiates a second CE episode and becomes a naked helium star.
Later the secondary explodes in a Type Ib/c (core-collapse) supernova, forming an NS 
 with a high natal kick ($229\kms$). Subsequently, an eccentric ($e=0.90$) BH-NS binary
with a moderate mass ratio forms after $10.5$ Myr of evolution since ZAMS.
After $2.7$ Gyr, the merger of the BH with the NS leads to a burst of gravitational 
      waves but without mass ejection or kilonova.

\begin{figure}[h]
\includegraphics[width=9.5cm, height=11cm]{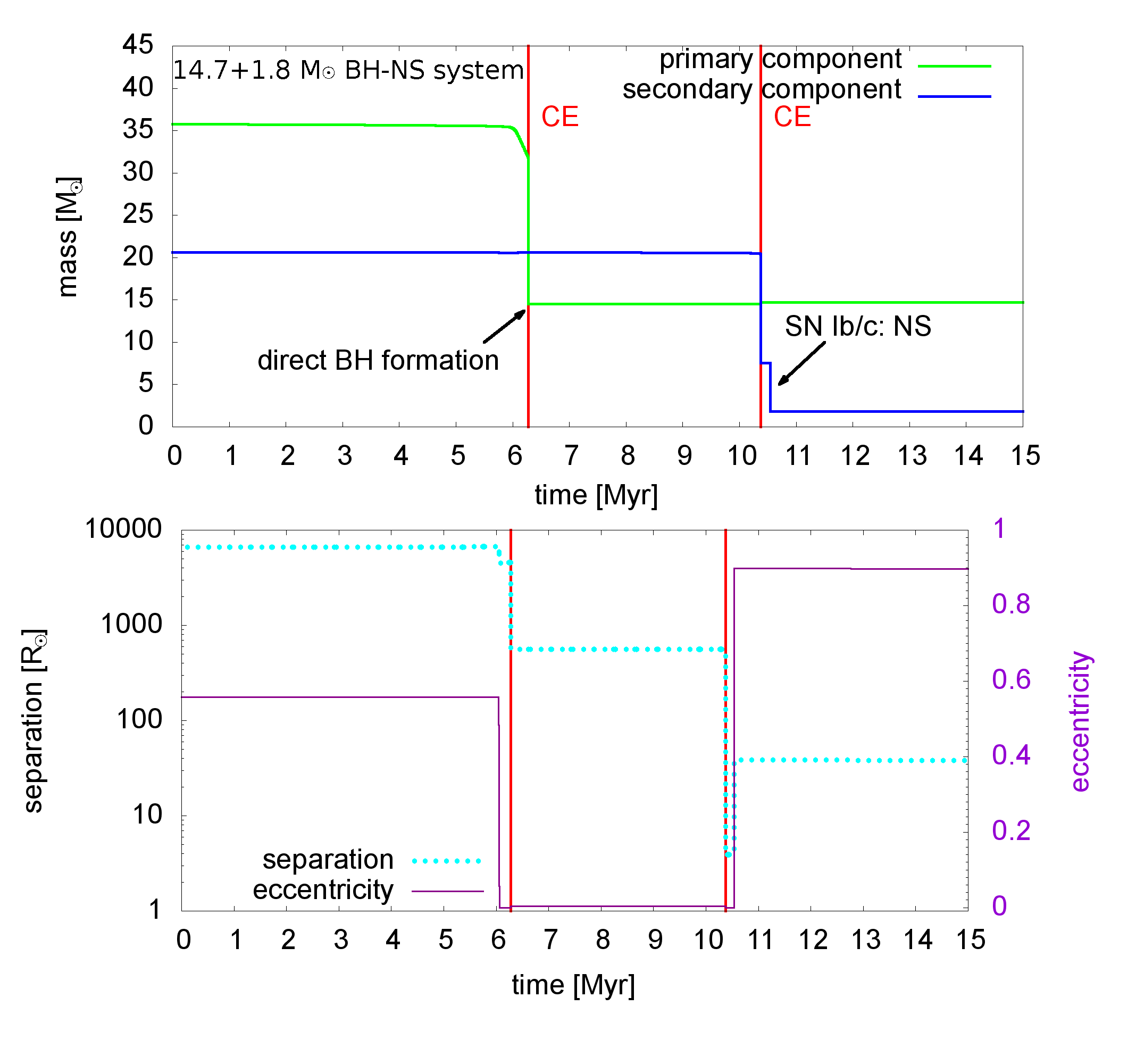}
\caption{Evolution of two massive stars that leads to the formation of the
moderate mass ratio ($q=0.12$) BH-NS ($14.7+1.8\msun$) merger (model M230). 
The top panel shows the mass evolution of the primary and secondary stars, while the
bottom panel shows the changes in the semimajor axis and eccentricity. Evolution 
from ZAMS to the BH-NS system formation takes $10.5$ Myr, and then it takes another 
$2.7$ Gyr for the two compact objects to merge and produce gravitational wave 
radiation that is potentially detectable by LIGO/Virgo. No mass ejection (no kilonova 
or short GRB) is expected in this case.} 
\label{fig.evol2}
\end{figure}

This typical $q=0.12$ system was formed from a ZAMS binary with comparable mass components: 
$q_{ZAMS}=0.58,$ and it evolved through two CE phases. Both components evolve along 
similar tracks. They form remnants shortly after the CE they initiated.
No tidal spin-up before the BH formation is expected: $a=557\rsun$ and $P_{\rm orb}=257$d. 
Both stars lose a significant fraction  ($\gtrsim 50\%$) of their mass in CE events, 
and the secondary component ejects some extra mass from the binary during a Type Ib/c 
supernova that forms a heavy NS. A high natal kick at the NS formation leads to the formation 
of a very eccentric ($e=0.9$) BH-NS binary that can merge within the Hubble time, despite 
its rather large separation ($a \sim 40\rsun$).

\subsection{Comparable mass BH-NS: $3.5+2.1\msun$ merger}
\label{3.5+2.1}

Comparable mass BH-NS systems, by construction, have  a high-mass NS and a  
low-mass BH, and typically one or both of the components have mass within the FMG. Therefore, such systems appear only within delayed supernova models. Below, we present 
the BH-NS merger with both components within the FMG ($3.5\msun$ BH $+2.1\msun$ NS; 
$q=0.6$) that has formed at metallicity $Z=0.0015$ in model M280.
Firstly, the ZAMS binary is formed with almost equal-mass components: $18.80\msun$ and 
      $18.66\msun$ at a moderately wide ($a=1100\rsun$) and eccentric 
      ($e=0.36$) orbit.
Later, when both components become core helium burning stars at a similar time,
 stable RLOF starts from the primary to the secondary.
As a result,  the primary component becomes a naked helium star after losing its H-rich envelope.
Then, the CE is initiated by the secondary component, which is still 
 a core helium burning star.
After the CE, both components are naked helium stars.
However, this state is not permanent: another stable RLOF takes place from the primary
 (evolved naked helium star) to the secondary component ($\sim 2.0\msun$ mass loss).
As a result the primary component explodes as a stripped supernova ($\sim 2.6\msun$ mass loss) 
with a $228\kms$ natal kick forming an NS, and the binary becomes eccentric ($e=0.24$).
The secondary component supernova explosion forms a BH and a $217\kms$ natal 
kick reduces system eccentricity ($e=0.06$).
Then, a close ($a=3.6\rsun$) BH-NS binary forms on an almost circular orbit after 
$\sim 12$ Myr evolution since ZAMS. After another $0.6$ Gyr, the BH-NS merger occurs.

\begin{figure}[h]
\includegraphics[width=9.5cm, height=11cm]{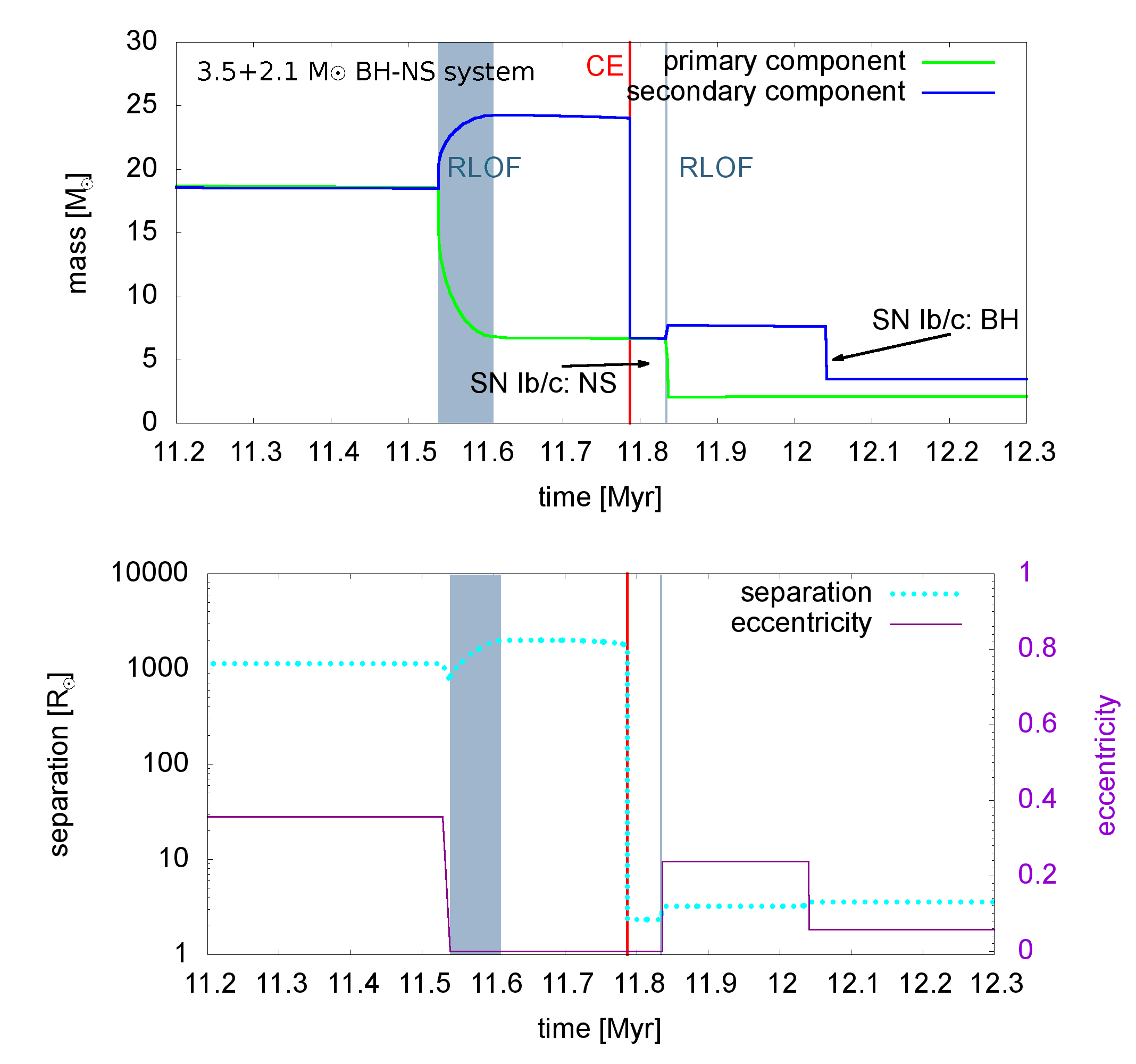}
\caption{Evolution of binary system that forms a $3.5+2.1$ $\msun$ BH-NS merger (mass ratio 
of $q=0.6$). We note that both compact objects are within the FMG and 
we expect that this merger will not lead to any mass ejection due to a high 
NS mass.}
\label{evol06}
\end{figure}

In this case, the NS forms $\sim 0.2$ Myr before the BH forms from the
primary star due to early mass ratio reversal between the primary and
the secondary (first stable RLOF).  Figure~\ref{evol06} shows the evolution
of the binary progenitor of this system in more detail.  The small
orbital separation prior to the BH formation ($a=2.3\rsun$, $P_{\rm orb}=0.11$d) 
causes the WR star progenitor of the BH to be spun up due to tidal 
interactions. The BH spin becomes very large: $a_{spin}=0.983$ in comparison 
to $a_{spin}=0.092$ expected from single stellar evolution without tidal 
interactions. Due to the high NS mass and its compactness, no mass ejection 
is expected even for the high spin and low mass of a BH.

\section{Mass ejection in BH-NS mergers}
\label{sec.mj}

Some BH-NS mergers may produce an electromagnetic counterpart.  For this to take place, 
the NS must disrupt outside the BH event horizon.  In this section we discuss
the amount of mass (if any) that is ejected during the BH-NS merger process. 

\begin{table}[h]
\caption{
Intrinsic fraction of BH-NS mergers ($z<1$) $\eta_2$ with any
 mass ejecta ($M_{\rm ej}>0.001\msun$) and with significant mass ejecta ($M_{\rm ej}~>~0.01\msun$).}
\centering
\begin {tabular}{l|cc}
\hline\hline
Model & $M_{\rm ej}>0.001\msun$ & $M_{\rm ej}>0.01\msun$ \\
\hline\hline
M230 &0.006 & 0.002\\
M233 & 0.063 & 0.022\\
M280 & 0.006 & $<10^{-4}$\\
M283 & 0.034 & $<10^{-4}$\\
M383 & 0.087 & 0.038\\
M483 & 0.030 & 0.004\\
\hline\hline
\multicolumn{3}{l}{Note: see eq.~\ref{BHNS_fit2}}\\
\end{tabular}
\label{ejecta_prc}
\end{table}

\begin{figure}[h]
\includegraphics[width=9.0cm, height=7cm]{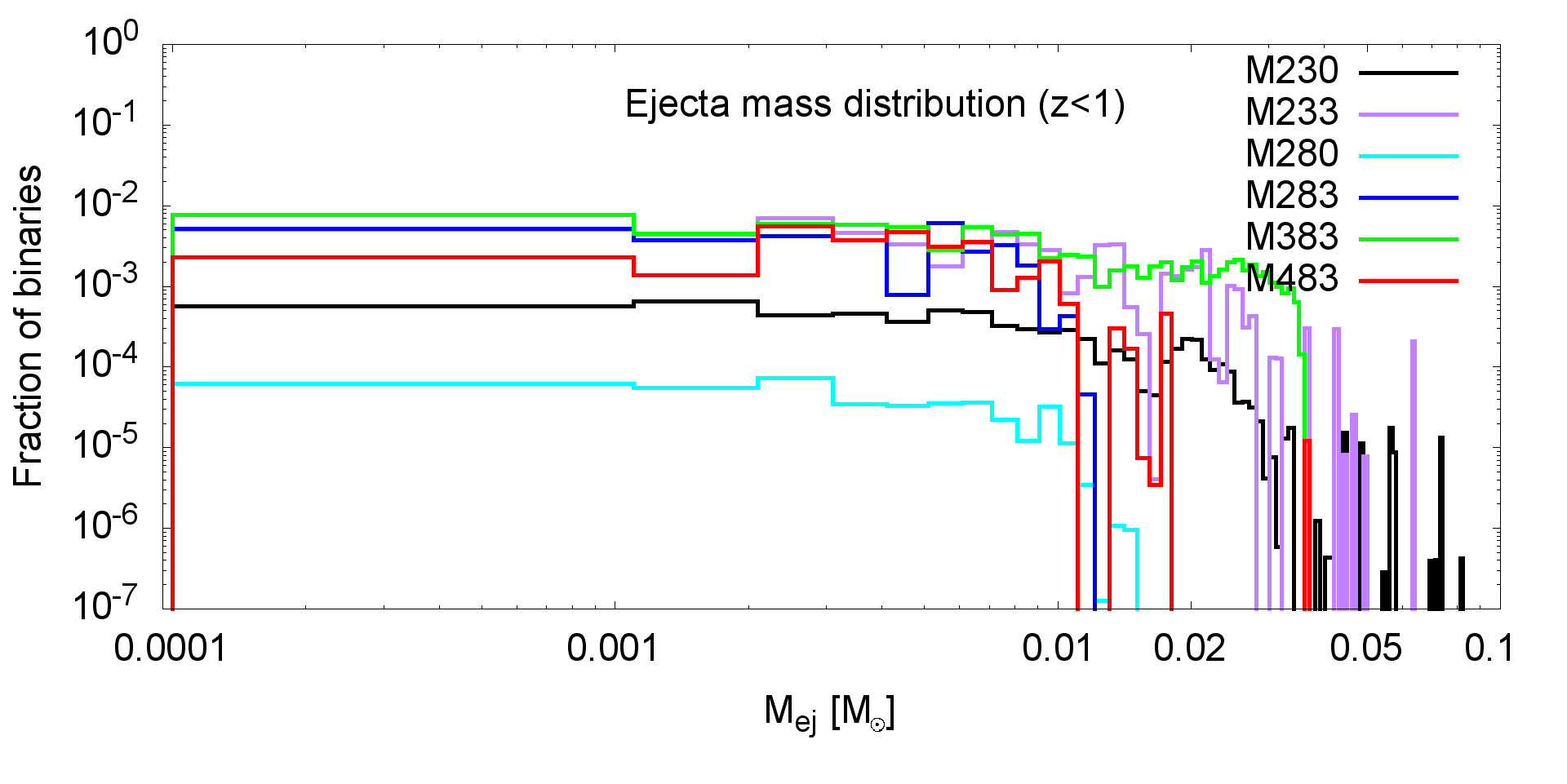}
\caption{Intrinsic distribution of ejecta mass in BH-NS mergers with~$z~<~1$.
The structures at $M_{ej} \gtrsim 0.03 \msun$ are shot noise.
}
\label{Mej_hist}
\end{figure}

\begin{figure}[h]
\includegraphics[width=9.5cm, height=7cm]{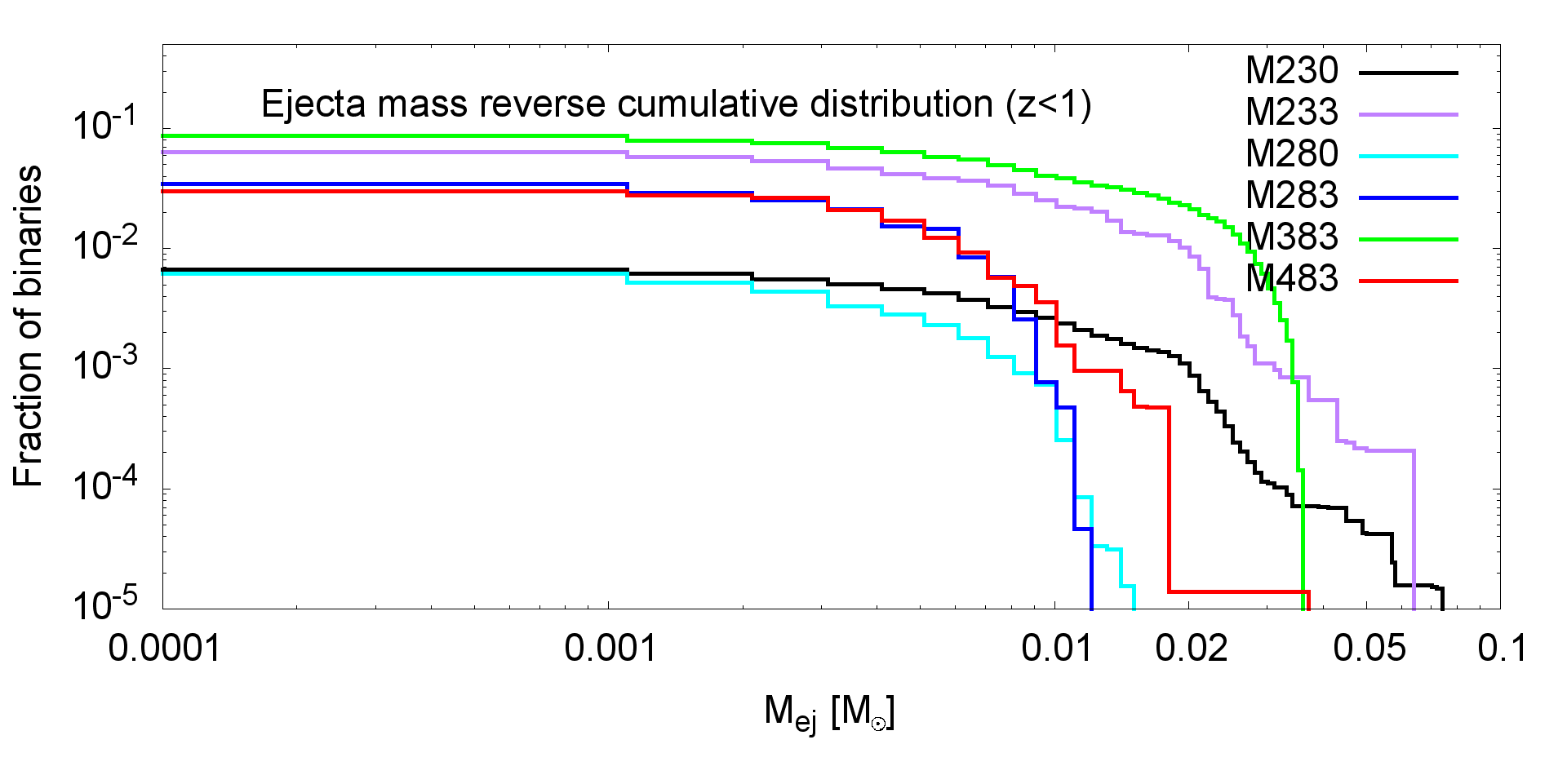}
\caption{Intrinsic reverse cumulative distribution of ejecta mass ($>M_{\rm ej}$) in 
BH-NS mergers with $z<1$.
}
\label{Mej}
\end{figure}

Table~\ref{ejecta_prc} shows the fraction of BH-NS mergers in our intrinsic 
population ($z<1$; eq.~\ref{BHNS_fit2}) with any mass ejecta ($M_{\rm ej}>0.001\msun$) 
and with significant mass ejecta ($M_{\rm ej}>0.01\msun$) for all our models. 
The corresponding distribution and reverse cumulative\footnote{By reverse
cumulative distribution we mean the fraction of systems with a higher value than 
the current argument} distributions are shown in Figures~\ref{Mej_hist} and ~\ref{Mej}, 
respectively.

The lowest ejecta mass found for BH-NS
mergers produced by our simulations and calculated with
equation~\ref{Mej_eq} is $M_{\rm ej}=0.0002\msun$, but we use $M_{\rm ej}=0.001\msun$ 
as the threshold for marking systems with {\em any} mass
ejection.  For our models that eject mass, the number of systems with
$M_{\rm ej}<0.001\msun$ is negligible.  The largest ejecta mass is
$M_{\rm ej} \sim 0.07\msun$, which is consistent with the results from
\cite{rosswog2005mergers,kyutoku2015dynamical, barbieri2020kilonova}.
The ejecta mass estimated for the first confirmed NS-NS merger,
GW170817: $\sim 0.0002-0.03\msun$~\citep{cote17,abbott2017estimating} is in the
range of our estimates.

For high BH spins (model M383; Geneva inefficient angular momentum transport), we obtain the 
maximum fraction of BH-NS mergers with any ejecta among our models: $\eta_2=0.087$
(see eq. \ref{BHNS_fit2}). This result is 
more pessimistic than earlier population synthesis predictions that showed the fraction of 
BH-NS mergers for rapidly spinning BHs with mass ejecta is as high as $0.4$~\citep{belczynski2008black}. 
This difference comes from the updated numerical simulations of ejecta mass, better constraints on the 
EoS, and an improved understanding of input stellar and binary physics. However, it should be noted that for low BH 
spins, our results are consistent with previous predictions that showed a $0.01$ fraction of BH-NS 
mergers with mass ejecta~\citep{belczynski2008black} . For comparison, our reference model with low 
BH spins (model M280: efficient MESA angular momentum transport) shows that the fraction of BH-NS mergers 
with any mass ejecta is comparable: $\eta_2=0.006$ (see Table~\ref{ejecta_prc}).

The rapid supernova engine models (M230 and M233) produce BH-NS mergers with low-mass NSs 
($\lesssim 2\msun$) and show the larger fraction of systems with mass ejecta than the
corresponding delayed supernova engine models (M280 and M283). In the delayed models, we also allow for the formation of heavy NSs ($>2\msun$), which are more compact 
than lower-mass NSs (see Fig.~\ref{fig.eos}) and thus are harder to disrupt.

BH spin plays a significant role in setting the size of the event horizon and this regulates 
whether NS disruption can produce any mass ejecta~\citep{foucart2018remnant,zappa2019black,
sedda2020dissecting}.
For rapidly rotating BHs (small event horizons), mass ejection is found in a small
fraction of BH-NS mergers (M383) and the fraction decreases for slowly spinning BHs 
(M230, M233, M280, and M283). There are still some BH-NS mergers that generate mass ejection 
for the (almost) nonspinning BH model (M483).

Table~\ref{MG_Mej} shows only those BH-NS mergers with any mass ejecta ($M_{ej}>0.001\msun$; 
see Table~\ref{ejecta_prc}) and we subdivide these to show the contribution of mergers in which 
none, one (BH or NS), or both components are within the FMG. These are, therefore, 
intrinsic fractions for the sub-population of BH-NS mergers (with any mass ejection) within $z<1$. 
Models with a rapid supernova engine (M230 and M233) have both merger components always outside the FMG. 
However, for our reference model (M280) and two other models with a delayed supernova engine (M283 
and M483) the fraction of BH-NS mergers with mass ejecta having both components in the FMG is 
very high $(\sim 0.9)$. 

Because binaries with both objects in the FMG correspond to more favorable mass ratios for 
tidal disruption outside the BH horizon, the fraction of binaries in the FMG that produce mass 
ejecta is high for delayed supernova engine models. 

In the most extreme case of our reference model (M280), we find that the fraction of BH-NS 
mergers with any mass ejecta that have both components within the FMG is $0.959$.
These BH-NS systems have evolved through spin-up of a WR star and BHs have a very large spin that 
allows for easy mass ejection ($a_{\rm spin}>0.9$). The lowest mass BHs (i.e., in the FMG) are 
naturally selected for systems with mass ejecta as they have the smallest event 
horizon. The heaviest NSs (i.e., in the FMG) are also favored  since in order to spin up the BH 
progenitor (a WR star), either an NS forms first (so it must be heavy; mass ratio reversal), or the 
NS progenitor is a WR star at the time when the BH progenitor is a WR star (so both stars evolve  
at almost similar timescales and must be similar in mass). Additionally, this model allows for 
fallback-moderated natal kicks, so heavy NSs receive smaller kicks and they have an increased 
chance of surviving in a binary after a supernova explosion than the lighter NSs. 

For the model with high initial BH spins (M383), the fraction of mergers with mass ejecta and 
both compact objects in the FMG is smaller ($0.365$), as specific processes (i.e., tidal spin up) 
are not the only way to produce mass ejecta.

\begin{table}[h]
\caption{Intrinsic BH-NS merger population with any mass ejecta ($M_{ej}>0.001\msun$, $z<1$).}
\centering
\begin {tabular}{l|ccccr}
\hline\hline
Model& $\rm{FMG_{both}}$ & $\rm{FMG_{BH}}$ & $\rm{FMG_{NS}}$ & $\rm{FMG_{none}}$ \\
\hline\hline
M230 & 0 & 0 & 0 & 1\\
M233 & 0 & 0 & 0 & 1\\
M280 & 0.959 & 0 & 0.005 & 0\\
M283 & 0.854 & 0.007 & 0.139 & 0\\ 
M383 & 0.365 & 0.004 & 0.016 & 0.614\\
M483 & 0.940 & 0 & 0.030 & 0.030\\
\hline\hline
\multicolumn{5}{l}{Note: see Table~\ref{ejecta_prc}. We present the fraction of mergers for which}\\
\multicolumn{5}{l}{both ($\rm{FMG_{both}}$) components, only a BH ($\rm{FMG_{BH}}$) component,}\\
\multicolumn{5}{l}{only an NS ($\rm{FMG_{NS}}$) component,}\\
\multicolumn{5}{l}{or no ($\rm{FMG_{none}}$) components are within the FMG.}\\
\end{tabular}
\label{MG_Mej}
\end{table}

\begin{table}[h]
\caption{
LIGO/Virgo detectable BH-NS mergers ($S/R>8$) with any mass ejecta 
$M_{ej}>0.001\msun$. We present the fraction  of mergers for which both 
($\rm{FMG_{both}}$) components, only a BH ($\rm{FMG_{BH}}$) component, only an NS ($\rm{FMG_{NS}}$)  component, or no ($\rm{FMG_{none}}$) components are within the FMG.
}
\centering
\begin {tabular}{l|ccccr}
\hline\hline
Model& $\rm{FMG_{both}}$ & $\rm{FMG_{BH}}$ & $\rm{FMG_{NS}}$ & $\rm{FMG_{none}}$ \\
\hline\hline
M230 & 0 & 0 & 0 & 1\\
M233 & 0 & 0 & 0 & 1\\
M280 & 1 & 0 & 0 & 0\\
M283 & 0.996 & 0.004 & 0 & 0\\ 
M383 & 0.431 & 0.075 & 0 & 0.494\\
M483 & 0.967 & 0 & $<10^{-3}$ & 0.032\\
\hline\hline
\multicolumn{5}{l}{Note: we present the fraction of mergers for which}\\
\multicolumn{5}{l}{both ($\rm{FMG_{both}}$) components, only a BH ($\rm{FMG_{BH}}$) component,}\\
\multicolumn{5}{l}{only an NS ($\rm{FMG_{NS}}$) component,}\\
\multicolumn{5}{l}{or no ($\rm{FMG_{none}}$) components are within the FMG.}\\
\end{tabular}
\label{MG_ligoMej}
\end{table}

In Table~\ref{MG_ligoMej} we list fractions of BH-NS mergers with compact objects 
in and out of the FMG, but only for systems that have any mass ejection 
($M_{ej}>0.001\msun$) and are detectable by LIGO/Virgo (S/R>8; see eq.~\ref{SNR}). 
These fractions are calculated using merger rates (see eq.~\ref{BHNS_fit4}).
This table shows how likely it is for a BH and an NS to be within the FMG for a LIGO/Virgo
BH-NS merger that may have detectable kilonova (see Sec.~\ref{sec.em}).

The results are qualitatively similar to those we found for the intrinsic population
of BH-NS mergers presented above in Table~\ref{MG_Mej}. The small to moderate 
quantitative differences arise from the fact that our adopted criteria for the LIGO/Virgo
detectability set the horizon redshift for (light) BH-NS mergers with mass ejection to 
only $z_{\rm hor} \sim 0.07$ (Table~\ref{MG_ligoMej}). For comparison, the horizon 
redshift for the  overall BH-NS merger LIGO/Virgo detectable population is 
$z_{\rm hor} \sim 0.1$. In contrast, the redshift cut that we employed in Table~\ref{MG_Mej} 
is $z=1$.

\section{Electromagnetic counterpart}
\label{sec.em}

BH-NS mergers in which the NS is disrupted outside the BH event horizon produce mass 
ejecta and accretion disks that may produce EM counterparts. In this section we 
discuss our results on kilonova emission associated with BH-NS mergers with mass ejection.
For LIGO/Virgo detectable events ($S/R>8$), we calculate kilonova luminosities for each 
BH-NS merger with some mass ejecta ($M_{\rm ej}>0.001\msun$) following the scheme outlined 
in Section~\ref{sec.mejkn}. For each merger event we employ the merger distance from  
our synthetic Universe (see Sec.~\ref{sec.cosmo}).

\begin{figure}[h]
\includegraphics[width=9cm, height=7cm]{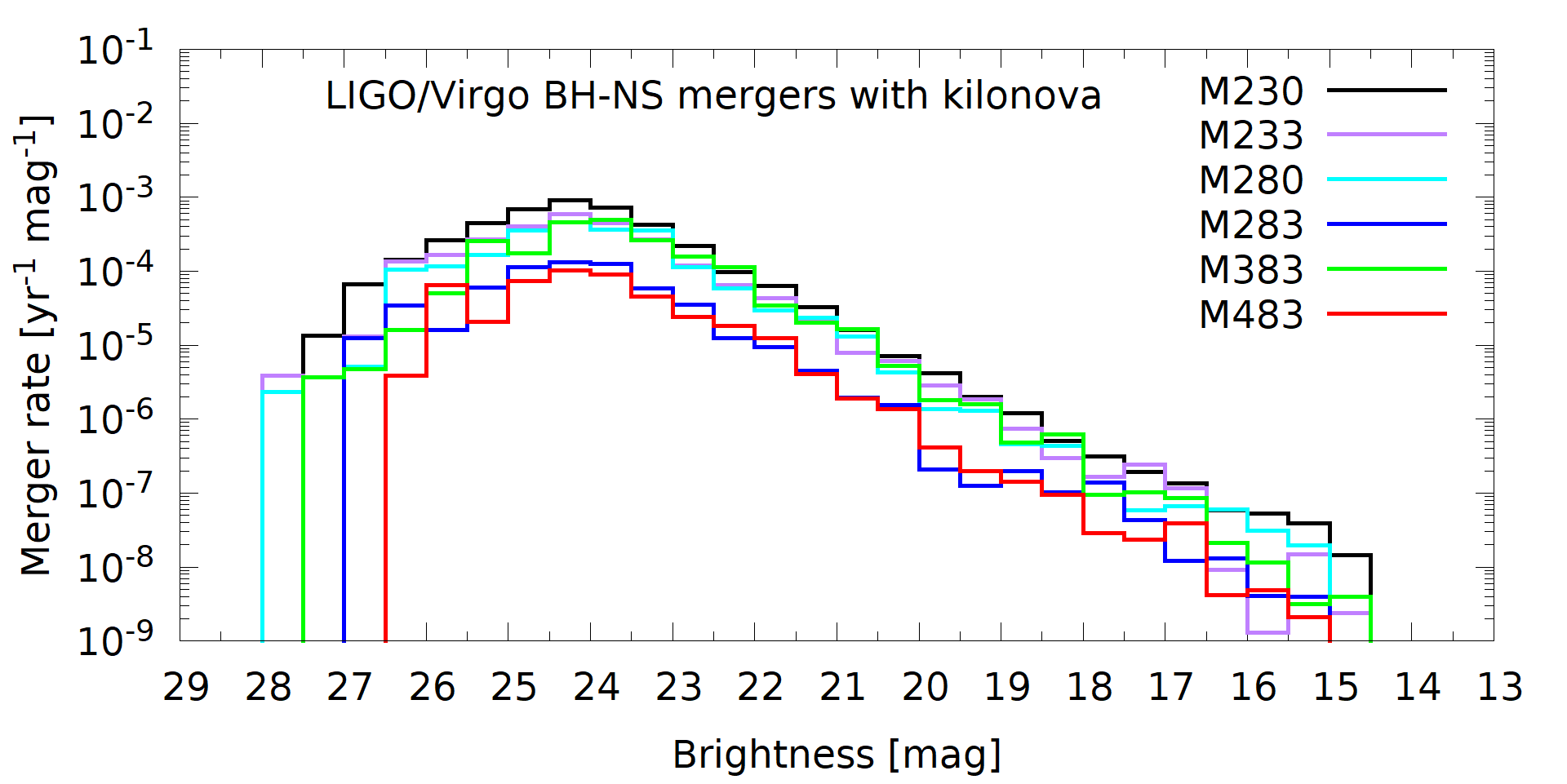}
\caption{
Kilonova brightness distribution for LIGO/Virgo detectable ($S/R>8$) BH-NS mergers.
}
\label{fig.kn}
\end{figure}

Figure~\ref{fig.kn} shows the distribution of brightness in o-band ($1260-1360{\rm nm}$) of BH-NS 
mergers with kilonova emission for all our models.  The distribution of kilonova brightness has a   very 
 large range ($M_{\rm kn} \sim 15-28$ mag), but dim kilonovae dominate ($M_{\rm kn} \sim 24$ mag).
 Since such events can be detected only at small redshifts
 (see Sec. \ref{sec.mj}), we do not take into account k-correction.
The large brightness range comes from the combination of distance range for 
BH-NS mergers that are detectable by LIGO/Virgo during O3 ($z \lesssim 0.07$; 
luminosity distance: $\lesssim 320$Mpc) and the fact that ejecta mass is found in 
a wide range of values ($M_{\rm ej}\sim 0.0002-0.07\msun$).  We find that  most 
kilonovae are dim because  most events are 
found at distances (average redshift: $z_{\rm ave}\sim 0.04$; average luminosity 
distance of $\sim 180$Mpc) close to the LIGO/Virgo O3 detection horizon for BH-NS 
mergers ($z \sim 0.07$; where most of the searched volume is located). Additionally, 
the ejecta mass found in our simulations is rather small; typically $M_{\rm ej}<0.03\msun$ 
(see Fig.~\ref{Mej_hist}).

\begin{table}[h]
\caption{Fraction of BH-NS mergers with detectable kilonova among LIGO/Virgo detectable ($S/R>8$) 
events for assumed exposure times. Kilonova detectability is assessed for three different instruments:
Subaru, the Canada-France-Hawaii Telescope (CFHT) and ATLAS. The same results expressed in terms of merger
rates can be found in the appendix; see Table~\ref{rate.kn}.}
\centering
\begin {tabular}{l|ccr}
\hline\hline
Model&Subaru&CFHT&ATLAS\\
\hline\hline
M230    &       0.004   &       0.002   &       $<10^{-5}$      \\
M233    &       0.067   &       0.032   &       $<10^{-4}$      \\
M280    &       0.004   &       0.002   &       $<10^{-5}$      \\
M283    &       0.021   &       0.010   &       $<10^{-4}$      \\
M383    &       0.064   &       0.038   &       $<10^{-4}$      \\
M483    &       0.018   &       0.009   &       $<10^{-4}$      \\
\hline\hline
\multicolumn{4}{l}{Note: see eq. \ref{BHNS_fit5}. }\\
\end{tabular}
\label{tab.kn}
\end{table}

Kilonova brightness varies with different model assumptions.
In Table~\ref{tab.kn}, for all our models, we show the fraction of BH-NS mergers with detectable kilonova among 
LIGO/Virgo detectable events ($S/R>8$; see eq. \ref{BHNS_fit5}).
We identify each simulated kilonova as detectable based on the detection thresholds of some typical optical instruments used in the search for kilonovae. 
We show results for small instruments such as ATLAS (maximum reach of $M_{\rm max}=19.5$mag), 
the medium-size Canada-France-Hawaii Telescope ($M_{\rm max}=24.1$mag) and the large Subaru 
telescope ($M_{\rm max}=26.0$mag).

We find that the fraction of potentially detectable kilonovae associated with BH-NS mergers 
is negligible for small instruments: in $10,000$ LIGO/Virgo detected BH-NS mergers,
only $\lesssim 1$ could be accompanied by a kilonova. For medium size instruments, the fraction 
of BH-NS mergers with associated detectable kilonovae varies from small ($\lesssim 4$ kilonova detections 
per $100$ LIGO/Virgo detections) to negligible ($2$ in $1000$). For large 
telescopes, a slightly larger fraction of kilonovae can be associated with LIGO/Virgo candidates: up to $\sim 7$ kilonovae per $100$ 
LIGO/Virgo BH-NS detections.  Much smaller fractions are possible; for example, this 
fraction drops to $4$ kilonovae per $1000$ LIGO/Virgo BH-NS detections for our reference model 
(M280) for an 8-meter class telescope.

\section{BH-NS merger rates}
\label{sec.rate}

In Table~\ref{rates} we list the BH-NS merger rate density ($R_{\rm d}$) obtained from our models 
at redshifts $z=0,\ 0.5$, and $\ 1$.  Merger rates increase with redshift (at least until redshift $z=2$) 
correlated with an increasing star formation rate. Merger rates for models with high NS/BH natal kicks (M233, 
M283, M383, and M483) are lower by more than $1$ order of magnitude than for models with natal 
kicks moderated by fallback (M230 and M280). 

\begin{figure}[h]
\includegraphics[width=9.0cm, height=12.0cm]{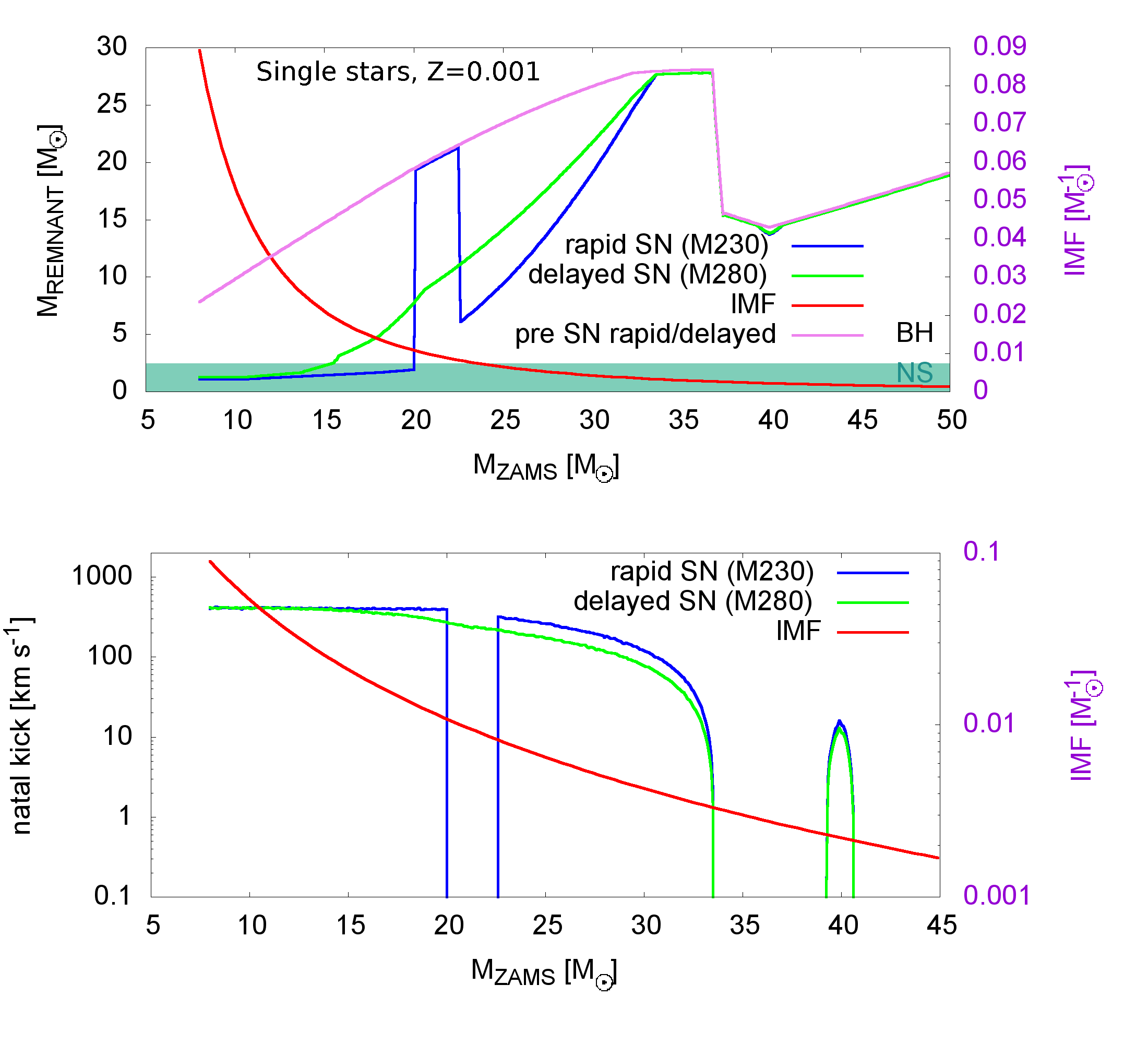}
\caption{
Top: initial (ZAMS) star mass vs. final ($M_{remnant}$) mass relation for single stars 
with a metallicity of $Z=0.001$ for two supernova engines: rapid employed (e.g., in 
model M230) and delayed employed (e.g., in model M280). The blue strip at the 
bottom marks the NS mass range adopted in our models ($M_{\rm NS}=1-2.5\msun$).
Bottom: fallback decreased natal kicks received by remnants (NS or BH) formed 
out of single stars ($Z=0.001$) for a given initial (ZAMS) mass for the two 
supernova engines. The shape of the adopted initial mass function (IMF) is also shown. 
}
\label{fig.final}
\end{figure}

Comparing between models with fixed physics but different supernova engines, our merger rates are a factor of $\sim 2$ larger in model M230 (rapid  supernova engine) compared to 
model M280 (delayed supernova engine). Both models employ  fallback decreased natal kicks. As shown 
in Figure~\ref{fig.final}, BHs begin forming just above $M_{ZAMS} \sim 20\msun$ for the rapid 
supernova engine. In particular, BHs that form from the lowest mass stars ($M_{ZAMS} \sim 20-23\msun$) 
have large enough masses ($M_{\rm BH} \sim 20\msun$) to form through direct collapse, 
and therefore these BHs receive no natal kicks. 
Binaries (potential BH-NS progenitors) with 
such stars always survive BH formation. On the other hand, for the delayed supernova engine, BHs begin 
forming just above $M_{ZAMS} \sim 15\msun$. In the mass range $M_{ZAMS} \sim 15-23\msun,$ these BHs 
form with low mass ($M_{\rm BH} \lesssim 10\msun$) and they tend to receive high natal 
kicks that can easily disrupt binary progenitors of BH-NS mergers. This leads to the BH-NS 
merger rate difference between models that employ rapid and delayed supernova engines. There are 
also some differences between the models for higher initial star masses in terms of BH 
masses and natal kicks, but they are not that important due to the steep IMF adopted in 
our calculations. The emergence of a peak in BH mass in the initial-final mass relation associated with the transition between NS and BH formation 
($M_{ZAMS} \sim 20\msun$) for the rapid supernova engine, and the continuous gradual BH mass increase 
with the initial star mass (at least for $M_{ZAMS} \lesssim 35\msun$) for the delayed supernova 
engine are the major reasons behind the rate difference. These particular features of BH mass 
dependence on initial star mass were explained in the context of underlying hydrodynamical 
simulations of supernova explosions that were used to create both supernova engine models
~\citep{fryer2012compact,belczynski2012missing}.

\begin{table}[h]
\caption{Intrinsic BH-NS merger rate density $R_{\rm d}$ [$\gpy$].}
\centering
\begin {tabular}{l|ccc}
\hline\hline
model & $R_{\rm d}$ ($z=0.0$) & $R_{\rm d}$ ($z=0.5$) & $R_{\rm d}$ ($z=1.0$) \\
\hline\hline
M230 &6.759 &13.189 &24.924  \\
M233 &0.328 &0.647 &1.196  \\ 
M280 &3.188 &6.166 &11.008  \\ 
M283 &0.245 &0.463 &0.859  \\ 
M383 &0.276 &0.474 &0.889  \\ 
M483 &0.214 &0.412 &0.732  \\ 
\hline\hline
\end{tabular}
\label{rates}
\end{table}

In previous sections we presented all of our results in terms of fractions of the BH-NS 
merger population. The rates listed in Table~\ref{rates} can be used to convert 
these fractions to merger rate densities. In some cases, a high merger rate can 
compensate for a low fraction of BH-NS mergers with desired properties, or viceversa.
For example, we estimated that the highest fraction of LIGO/Virgo detected BH-NS 
mergers will be associated with kilonovae for models M233 and M383, while other 
models provide $\sim3-16$ times smaller fractions (see Table~\ref{tab.kn}). However, 
models M233 and M383 have much smaller (by $\lesssim 10$)  BH-NS merger rate 
densities than models M230 and M280. 

The merger rate density of BH-NS systems was estimated from the O1 and O2 LIGO/Virgo observations to 
be $<610\gpy$ \citep{LIGO2019b}. Using one BH-NS merger candidate from  the O3 LIGO/Virgo 
observations (GW190814), \cite{belczynski2019evolutionary} estimated the rate density of 
BH-NS mergers to be in the range of $1.6-60\gpy$. However, it was noted by the LIGO/Virgo 
Collaboration that this may be a BH-BH merger~\citep{gw190814}. More recently, based on two 
BH-NS merger candidates (GW200105 and GW200115), the LIGO/Virgo Collaboration estimated the BH-NS 
merger rate to be in the range of $7.4-320\gpy$ ($90\%$ credible level; ~\cite{LIGO2021c}). 

Our calculated BH-NS merger rate densities among the tested models are: $0.2-6.7\gpy$ ($z=0$), 
$0.4-13.2\gpy$ ($z=0.5$), and $0.7-24.9\gpy$ ($z=1$). These predicted rates are somewhat small 
compared to the most recent LIGO/Virgo estimates, especially since  LIGO/Virgo 
can currently detect BH-NS mergers only to a redshift of $z \sim 0.1$. There are a number of factors that 
have a significant impact on merger rates and that we have not tested here. The star formation rate 
and associated cosmic metallicity evolution have nontrivial impacts on derived merger
rates~\citep{Chruslinska2019a,Chruslinska2019b,Tang2020,Boco2021,Santoliquido2021}.   
New developments to understand the physics of the formation of merging double compact objects 
may also lead to significant changes in merger rates. For example, in this study we limited 
our modeling to the formation of BH-NS mergers through the CE isolated binary 
evolution channel. However, a non-CE channel of isolated binary evolution may produce BH-NS 
mergers as well, altering the merger rates and BH-NS properties~\citep{Heuvel2017,
Stevenson2019,Neijssel2019,Shao2021,Olejak2021}. We plan to dedicate a separate study to 
assessing the impact of these processes on BH-NS merger populations.

\section{Conclusions}

We performed a suite of binary evolution calculations to model a population of BH-NS 
mergers. Such mergers are gravitational wave sources and can potentially also 
produce EM emission (e.g., kilonova, short GRB). They are also interesting in the 
context of existence (or inexistence) of the FMG, the dearth of compact 
objects (whether they are NSs or BHs) in the mass range $\sim 2-5\msun$. At present, 
it seems that EM observations are beginning to place some compact objects in this mass 
range \citep{cromartie2019very,thompson2019noninteracting}, and the same is found for 
GW LIGO/Virgo detections \citep{gw190814}. Our study may be summarized as follows: 

\begin{enumerate}

\item We predict that only a very small fraction of BH-NS systems will be accompanied 
by mass ejection during the merger process. The fraction of such systems within the overall 
BH-NS merger population depends sensitively ($0.6-9\%$; see Table~\ref{ejecta_prc}) on 
 uncertain input physics.  
For our best guesses of input physics (model M280), we predict that  $\lesssim 1\%$
of BH-NS mergers are accompanied by mass ejection and by kilonovae bright enough to be 
detectable with a large $8$-meter class telescope. This result was obtained for a 
specific MPA1$EoS$ (see Fig.~\ref{fig.eos}) and for rather optimistic 
assumptions on kilonova detectability (see Sec.~\ref{presentation}).
Since the mass ejection (and EM 
detectability) depends sensitively on BH spin, the future detections (or lack thereof) 
of BH-NS mergers in EM may distinguish among several existing models of angular momentum 
transport in massive stars that sets the BH natal spin
~\citep{Bavera2020,belczynski2019evolutionary}.

\item If more compact objects within the FMG are found, this would place 
limits such as those assumed in the ``delayed'' explosion scenario from~\cite{fryer2012compact}, 
with weaker explosions with material accretion (e.g., fallback) after the launch of the 
explosion. A qualitatively similar result is also obtained by~\cite{Zevin2020}. In our 
scenarios with a delayed supernova engine, we find that a significant fraction of BH-NS 
mergers may host a FMG compact object: $\gtrsim 30\%$ (models M280, M283, M383, 
and M483; see Table~\ref{MG_intrinsic}). 

\item Our models indicate that extreme BH-NS mass ratio systems may form in 
classical isolated binary evolution. For example, it is possible for systems to form with a 
mass ratio (NS mass to BH mass) as low as $q=0.02$ with $\sim 1\msun$ NS and 
$\sim 50\msun$ BH. However, such systems are only a very small part of the overall BH-NS 
merger population, and typical BH-NS mergers are expected to form with mass ratios of 
$q=0.1-0.2$. 

\end{enumerate}

Our current binary evolution models with BH-NS mergers (and also with BH-BH and NS-NS) 
mergers are publicly available for further exploration at 
\footnote{\url{https://syntheticuniverse.org}}.

\section{Acknowledgements}
We would like to thank the anonymous referee for their very useful and detailed comments and
suggestions that helped to improve our manuscript. 
We would like to thank  Ben Farr, Martyna Chruslinska, Coen Neijssel, Zoheyr Doctor
and all the participants of the $"$Astrophysics of LIGO/Virgo sources in O3 era $"$
conference who suggested research improvements.  
ROS would like to thank Erika Holmbeck for her helpful discussions.
KB and PD acknowledge support from the Polish National Science Center (NCN) grant  
Maestro (2018/30/A/ST9/00050). TB was supported by the FNP grant Team/2016-3/19
and NCN grant UMO-2017/26/M/ST9/00978. ROS acknowledges support from NSF AST-1909534.


\bibliographystyle{aa}
\bibliography{ms}

\begin{appendix}
\section{Merger rates}
\label{sec.appendix}

In Table~\ref{rate.MG_intrinsic} we list the intrinsic merger rate density of BH-NS systems 
with compact objects in the FMG for all models and for the local Universe 
($z\sim0$). Data in this table are the combination of Table~\ref{MG_intrinsic} 
and the first column of Table~\ref{rates}. In other words, it is assumed that fractions of 
BH-NS mergers with various combinations of compact objects in the FMG
(Tab.~\ref{MG_intrinsic}) are constant in the redshift range $z=0-1$. 

In Table~\ref{rate.extreme} we list the merger rate density of LIGO/Virgo detectable BH-NS 
systems with extreme and small mass ratios for all the models. Data in this Table are 
the combination of Table~\ref{tab.extreme} and the first column of Table~\ref{rates} (local 
Universe $z\sim0$). 

In Table~\ref{rate.kn} we list the merger rate density of LIGO/Virgo detectable BH-NS 
systems with detectable kilonovae for all the models. Data in this Table are the combination 
of Table~\ref{tab.kn} and the first column of Table~\ref{rates} (local Universe $z\sim0$). 
 
\begin{table}[h]
\caption{
Intrinsic merger rate density [$\gpy$] of BH-NS systems in which both ($\rm{FMG_{both}}$) components, 
only a BH ($\rm{FMG_{BH}}$) component, only an NS ($\rm{FMG_{NS}}$) component, or no ($\rm{FMG_{none}}$) components are within the FMG.}
\centering
\begin {tabular}{l|ccccr}
\hline\hline
Model&$ \rm{FMG_{both}}$ & $\rm{FMG_{BH}}$ & $\rm{FMG_{NS}}$ & $\rm{FMG_{none}}$ \\
\hline\hline
M230    &       0.000   &       0.000   &       0.000   &       6.759   \\
M233    &       0.000   &       0.000   &       0.000   &       0.328   \\
M280    &       0.115   &       0.061   &       0.756   &       2.257   \\
M283    &       0.048   &       0.027   &       0.025   &       0.145   \\
M383    &       0.055   &       0.027   &       0.033   &       0.161   \\
M483    &       0.045   &       0.023   &       0.022   &       0.123   \\
\hline\hline
\end{tabular}
\label{rate.MG_intrinsic}
\end{table}

\begin{table}[h]
\caption{Merger rate density [$\gpy$] of LIGO/Virgo detectable ($S/R>8$) 
BH-NS systems with extreme and small mass ratios.}
\centering
\begin {tabular}{l|cccc}
\hline\hline
Model & q<1/50 & q<1/30 & q<1/20 & q<1/10 \\
\hline\hline
M230    &       0.000   &       0.002   &       0.011   &       0.224   \\
M233    &       0.000   &       0.001   &       0.010   &       0.035   \\
M280    &       0.000   &       0.000   &       0.007   &       0.066   \\
M283    &       0.000   &       0.004   &       0.007   &       0.029   \\
M383    &       0.000   &       0.001   &       0.004   &       0.036   \\
M483    &       0.000   &       0.001   &       0.003   &       0.035   \\
\hline\hline
\end{tabular}
\label{rate.extreme}
\end{table}

\begin{table}[h]
\caption{
Merger rate density [$\gpy$] of BH-NS systems with detectable 
kilonova among LIGO/Virgo detectable ($S/R>8$) events for our assumed exposure times.}
\centering
\begin {tabular}{l|ccc}
\hline\hline
Model& Subaru & CFHT & ATLAS \\
\hline\hline
M230    &       0.0270  &       0.0135  &       $<10^{-5}$      \\
M233    &       0.0220  &       0.0105  &       $<10^{-4}$      \\
M280    &       0.0128  &       0.0064  &       $<10^{-5}$      \\
M283    &       0.0051  &       0.0025  &       $<10^{-3}$      \\
M383    &       0.0177  &       0.0105  &       $<10^{-4}$      \\
M483    &       0.0039  &       0.0019  &       $<10^{-4}$      \\
\hline\hline
\multicolumn{4}{l}{The kilonova detectability is assessed for Subaru,}\\
\multicolumn{4}{l}{the Canada-France-Hawaii Telescope (CFHT)}\\
\multicolumn{4}{l}{and ATLAS.}\\
\end{tabular}
\label{rate.kn}
\end{table}

\end{appendix}

\end{document}